\documentclass[%
 reprint,
 amsmath,amssymb,
 aps,
]{revtex4-2}

\usepackage[utf8]{inputenc} 
\usepackage{graphicx}

\usepackage{csquotes}

\usepackage{fancyhdr}
\usepackage{float}
\usepackage{amssymb}
\usepackage{upgreek}
\usepackage{amsmath}
\usepackage{indentfirst}

\usepackage{dsfont}

\usepackage[english]{babel}
\usepackage[usenames, dvipsnames]{color}
\usepackage{soul}
\usepackage{multirow}
\usepackage{ragged2e}
\usepackage{mdframed}
\usepackage{gensymb}
\usepackage{physics}
\usepackage{comment}
\usepackage{wrapfig}
\usepackage{emptypage}

\usepackage{pgfplots}
\usepgfplotslibrary{fillbetween,groupplots,external}
\pgfplotsset{compat=newest}
\pgfplotsset{plot coordinates/math parser=false}
\newlength\figureheight
\newlength\figurewidth 

\pgfplotsset{aspect ratio/.code args={#1:#2}{%
		\def\axisdefaultwidth{#1 cm}%
		\def\axisdefaultheight{#2 cm}},
	compat=newest
}

\usepackage{bm}

\usepackage[colorlinks = true, citecolor = blue, linkcolor = blue, urlcolor=blue]{hyperref}

\usepackage{xcolor}

\usepackage{tikz}
\usepackage{tikz-3dplot}
\usetikzlibrary{3d,spy,external,matrix,plotmarks,pgfplots.groupplots,calc,decorations.pathreplacing,decorations.pathmorphing,arrows, decorations.markings, fit, backgrounds, arrows.meta, positioning}

\definecolor{myyellow}{RGB}{240,217,1}
\definecolor{mygreen}{RGB}{143,188,103}
\definecolor{myred}{RGB}{234,38,40}
\definecolor{myblue}{RGB}{53,101,167}
\definecolor{mygray}{RGB}{192,192,192}

\pgfdeclaredecoration{newarrow}{initial}{
    \state{initial}[%
        switch if less than=\pgfdecoratedpathlength/1 to final, 
        width=\pgfdecoratedpathlength/2, 
        next state=final
        ]
    {%
        \pgfsetfillcolor{white}
        \pgfsetstrokecolor{black}
        \pgfsetmiterjoin \pgfsetmiterlimit{12} 
        \pgfmathsetmacro\onethird{\pgfdecoratedpathlength/2*1.5}
        \pgfmathsetmacro\arrowhead{\pgfdecoratedpathlength/7}
        \pgfmathsetmacro\arrowspread{\arrowhead/2}
        \pgfmoveto{\pgfpoint{14pt}{0pt}} 
        \pgflineto{\pgfpoint{\onethird}{\arrowspread}}
        \pgflineto{\pgfpoint{\onethird}{\arrowhead}}
        \pgflineto{\pgfpoint{\pgfdecoratedpathlength-3pt}{0pt}}
        \pgflineto{\pgfpoint{\onethird}{-\arrowhead}}
        \pgflineto{\pgfpoint{\onethird}{-\arrowspread}}     
        \pgfclosepath
        \pgfmoveto{\pgfpoint{\pgfdecoratedpathlength}{0pt}}
    }
    \state{final}
    {%
        \pgfpathlineto{\pgfpointdecoratedpathlast}
    }
}
\tikzset{
    custarr/.style={
        decorate, decoration={name=newarrow}%
    }
}

\pgfdeclaredecoration{newarrow2}{initial}{
    \state{initial}[%
        switch if less than=\pgfdecoratedpathlength/1 to final, 
        width=\pgfdecoratedpathlength/2, 
        next state=final
        ]
    {%
        \pgfsetfillcolor{white}
        \pgfsetstrokecolor{black}
        \pgfsetmiterjoin \pgfsetmiterlimit{12} 
        \pgfmathsetmacro\onethird{\pgfdecoratedpathlength/2*1.9}
        \pgfmathsetmacro\arrowhead{10pt}
        \pgfmathsetmacro\arrowspread{\arrowhead/2}
        \pgfmoveto{\pgfpoint{-10pt}{\arrowspread}} 
        \pgflineto{\pgfpoint{\onethird}{\arrowspread}}
        \pgflineto{\pgfpoint{\onethird}{\arrowhead}}
        \pgflineto{\pgfpoint{\pgfdecoratedpathlength+15pt}{0pt}}
        \pgflineto{\pgfpoint{\onethird}{-\arrowhead}}
        \pgflineto{\pgfpoint{\onethird}{-\arrowspread}} 
        \pgflineto{\pgfpoint{-10pt}{-\arrowspread}}
        \pgfclosepath
        \pgfmoveto{\pgfpoint{\pgfdecoratedpathlength}{0pt}}
    }
    \state{final}
    {%
        \pgfpathlineto{\pgfpointdecoratedpathlast}
    }
}
\tikzset{
    custarr2/.style={
        decorate, decoration={name=newarrow2}%
    }
}

\newmdenv[outerlinewidth=1,outerlinecolor=grey,frametitle=Démonstration,innertopmargin=10pt,leftmargin=0,rightmargin=0]{proof}

\newmdenv[outerlinewidth=1,outerlinecolor=grey,innertopmargin=10pt,leftmargin=0,rightmargin=0]{customtitle}

\newmdenv[linewidth=1,linecolor=grey,innertopmargin=10pt,leftmargin=40,rightmargin=40]{see}

\let\oldhat\hat

\renewcommand{\hat}[1]{\oldhat{\bm{\mathbf{#1}}}}

\usepackage[T1]{fontenc} 

\pgfdeclaredecoration{complete sines}{initial}
{
	\state{initial}[
	width=+0pt,
	next state=sine,
	persistent precomputation={\pgfmathsetmacro\matchinglength{
			\pgfdecoratedinputsegmentlength / int(\pgfdecoratedinputsegmentlength/\pgfdecorationsegmentlength)}
		\setlength{\pgfdecorationsegmentlength}{\matchinglength pt}
	}] {}
	\state{sine}[width=\pgfdecorationsegmentlength]{
		\pgfpathsine{\pgfpoint{0.25\pgfdecorationsegmentlength}{0.5\pgfdecorationsegmentamplitude}}
		\pgfpathcosine{\pgfpoint{0.25\pgfdecorationsegmentlength}{-0.5\pgfdecorationsegmentamplitude}}
		\pgfpathsine{\pgfpoint{0.25\pgfdecorationsegmentlength}{-0.5\pgfdecorationsegmentamplitude}}
		\pgfpathcosine{\pgfpoint{0.25\pgfdecorationsegmentlength}{0.5\pgfdecorationsegmentamplitude}}
	}
	\state{final}{}
}

\tikzset{
	photon/.style={decorate, decoration={
			complete sines,
			segment length=2.5mm,
			amplitude=1.5mm
		}, draw=black},
	particle/.style={draw=black, postaction={decorate},
		decoration={markings,mark=at position .5 with {\arrow[draw=black]{>}}}},
	antiparticle/.style={draw=black, postaction={decorate},
		decoration={markings,mark=at position .5 with {\arrow[draw=black]{<}}}},
}

\usepackage{graphicx}
\usepackage{dcolumn}

\usepackage{textcomp}
\usepackage{eurosym}
\def\plusheight{-\the\dimexpr\fontdimen22\textfont2\relax}

\usepackage{nicematrix}
\NiceMatrixOptions{
	code-for-first-row = \color{gray} \scriptstyle ,
	code-for-last-row = \color{gray} \scriptstyle,
	code-for-first-col = \color{gray} \scriptstyle,
	code-for-last-col = \color{gray} \scriptstyle
}

\usepackage{bm}

\usepackage{xargs}
\usepackage{ifthen}
\newcommandx{\figref}[2][2=emptyarg]{\ifthenelse{\equal{#2}{emptyarg}}{Fig.~\ref{#1}}{Fig.~\hyperref[#1]{\ref*{#1}\textcolor{blue}{({#2})}}}}

\newcommand{\ti}{\tilde{I}}
\newcommand{\tx}{\tilde{X}}
\newcommand{\ty}{\tilde{Y}}
\newcommand{\tz}{\tilde{Z}}

\newcommand{\tm}{\tilde{M}}

\begin{document}

\title{Operator-space fragmentation and integrability in Pauli-Lindblad models}
\author{Dawid Paszko}
\author{Christopher J. Turner}
\author{Dominic C. Rose}
\author{Arijeet Pal}
\affiliation{Department of Physics and Astronomy, University College London, Gower Street, London WC1E 6BT, United Kingdom}

\begin{abstract}
The Lindblad equation for open quantum systems is central to our understanding of coherence and entanglement in the presence of Markovian dissipation. In closed quantum systems Hilbert-space fragmentation is an effective mechanism for slowing decoherence in the presence of constrained interactions. We develop a general mechanism for operator-space fragmentation of mixed states, undergoing Lindbladian evolution generated by frustration-free Hamiltonians and Pauli-string jump operators. The interplay of generator algebras of dissipative and unitary dynamics leads to a hierarchical partitioning of operator and real space into dynamically disconnected subspaces, which we elucidate using the bond and commutant algebras of superoperators. This fragmentation fundamentally constrains information spreading in open systems and provides new mechanisms to control highly entangled quantum states and dynamics. Our approach yields two key advances. Firstly, we introduce frustration graphs in operator space as a compact representation to construct effective non-Hermitian Hamiltonians in individual fragments and diagnose their free-fermion solvability. Secondly, using these methods we uncover a range of universal dynamical regimes in Pauli-Lindblad models, exhibiting symmetry enriched quantum chaos and integrability in operator-space fragments. Furthermore, we show dissipation-driven phase transitions corresponding to exceptional points in the Lindbladian spectrum whose signatures are captured by spectral statistics and operator dynamics. These results establish operator-space fragmentation as a fundamental principle for open quantum systems, with immediate implications for quantum error correction, where protected subspaces could emerge naturally from fragmentation. Our framework provides a systematic approach to discover and characterize novel non-equilibrium phases in open quantum many-body systems.
\end{abstract}

\maketitle


\section{Introduction}
\label{sec:intro}
Noise induced errors in quantum systems are ubiquitous. They lead to the depletion of coherence and entanglement which are fundamental resources in quantum information processing~\cite{Chuang1995}. The mechanisms to protect information encoded in quantum many-body states from noise can be utilized for quantum computing and stabilizing macroscopically entangled states~\cite{Shor1995, Dur_Briegel_PRL2004}. Entangled quantum states, acting as quantum error correcting codes, can be leveraged to reverse the impact of noise to process quantum information and control the propagation of errors~\cite{Calderbank1996, Steane1996, gottesman1997stabilizer, Terhal2015}. Despite the sensitivity of coherence to a noisy environment, controlled dynamics of entanglement in error correcting codes provide a physical pathway for robust information processing~\cite{kitaev1997quantum, bluvstein2024}. New regimes of entanglement dynamics emerge when interacting qubits are coupled to an environment, providing an exciting opportunity to stabilize entanglement and information in mixed states for a long time~\cite{Verstraete2009, PhysRevA.78.042307, Diehl2011, Paszko2024, Fan2024, lee2025SPT}. 

There are several experimental platforms such as circuit QED~\cite{Fitzpatrick2017}, photonics~\cite{bloch2022strongly, clark2020Laughlin}, and ultra-cold atoms~\cite{landig2016} with controllable dissipation which realize long-range correlated mixed states. These systems can be harnessed to stabilize topological order and protect information propagation in open quantum systems~\cite{Sohal2025, Ma2025, Chirame2025}. For a Markovian environment, the dynamics of the system is described by a Lindbladian superoperator acting in the space of density operators. The structure of the spectrum and eigenoperators of the Lindbladian encode the dynamical and steady-state properties of the system~\cite{Macieszczak2016, Ashida2018, garrahan2018aspects, buvca2019non, Sa2023, Kawabata2023, Kawabata2023b, guo2025designing}. In analogy with the behaviour of Hamiltonian operators governing unitary dynamics, the properties of the Lindbladian operators exhibit localization~\cite{deAlbornoz2024, Thompson2024, Liu2024} and fragmentation~\cite{PhysRevE.102.062210, PhysRevResearch.5.043239} opening the avenue for new classes of universal quantum many-body dynamics. Recently, applying methods and ideas from integrable systems have opened new directions in the study of Lindbladians and raises important questions about the universality of dynamics in open quantum systems~\cite{PhysRevLett.122.010401, PhysRevLett.117.137202, SciPostPhys.8.3.044, deLeeuw2021}.  

Dissipation plays a vital role in the dynamics of information and stability of memory~\cite{Harrington2022}. Memory effects arising from ergodicity-breaking due to many-body localization~\cite{BASKO20061126,PhysRevB.82.174411,PhysRevLett.111.127201,PhysRevB.90.174202} and scarring~\cite{Bernien2017,Turner2018, Moudgalya_2022} are unstable to coupling to a bath~\cite{Levi2016, Medvedyeva2016, Wang2024}. 
Solvable limits of interacting dissipative systems shed light on the robustness of non-ergodic phenomena and characterize the stability of highly-entangled states. In isolated systems the presence of Hilbert-space fragmentation (HSF) impedes thermalisation~\cite{PhysRevX.10.011047, PhysRevB.101.174204}, where the dynamics exhibit a range of behaviour from coherent oscillation between scarred eigenstates to anomalous hydrodynamics~\cite{PhysRevB.98.155134,PhysRevB.98.235156,PhysRevB.98.235155, Lee2021, Richter2022}. A form of HSF in open quantum systems has recently been proposed as a mechanism to stabilize highly entangled steady states~\cite{PhysRevResearch.5.043239, Paszko2024, Li2025}. Integrability of the Lindbladian superoperator can give rise to the operator space fragmentation (OSF) which influences density matrices. The algebraic structure of integrable symmetries of one-dimensional Hamiltonians provides a natural choice of jump operators in Lindbladian models for OSF. A unifying framework for OSF in open quantum systems and the role of symmetries in generating universal dynamics in Lindbladian models are open questions of interest.

In this work we explore a class of open quantum systems described by Pauli-Lindblad models which are salient for simulating noisy dynamics in quantum error correcting codes~\cite{vanDenBerg2023Qerror}. In these models the unitary evolution and the jump operators describing the effect of dissipation on the system are given by Pauli string operators. The Lindbladian dynamics evolves an initial density matrix over the space of operators which can lead to fully mixed steady states from the Gibbs ensemble~\cite{brandao2019finite, deLeeuw2024, rouze2024efficient, ding2025efficient}. Alternatively, the operator space can be fragmented due to the interplay of interaction and dissipation terms in the Lindbladian.
A form of fragmentation discussed in Ref.~\cite{PhysRevE.102.062210} was associated with the integrability of the Lindbladian superoperator, which has been generalized to $U(1)$ and $SU(N)$ symmetries~\cite{PhysRevResearch.5.043239, Li2025}. Operator space fragmentation has also recently been studied in the unitary dynamics of random Floquet circuits and its relationship to ergodicity~\cite{kovacs2024}. Our work generalises this notion of fragmentation to a class of models which can exhibit a wide range of dynamical behaviour given by chaotic spectral properties, integrability, and non-hermitian criticality at exceptional points.

We formulate a theory of operator-space fragmentation in open systems using bond and commutant algebras generated by the Hamiltonian and dissipative operators. This formalism was previously used to describe HSF in closed systems~\cite{PhysRevX.12.011050} and its aforementioned extension to open systems~\cite{PhysRevResearch.5.043239}. The bond algebra contains individual terms of the Hamiltonian and jump operators which capture the coupling to the Makovian bath, while the commutant contains all the operators that commute with the bond algebra. We generalize them to the algebra of superoperators and develop a theory of operator space fragmentation for Pauli-Lindblad models and elucidate the fragmentation properties of these algebras.
As an illustration, we provide an entire family of Pauli-Lindbladian models whose operator space exhibits a hierarchy of fragmentation. In the first step, the operator space fragments into regions which are dynamically disconnected from each other, which is dubbed as \textit{subspace fragmentation}. Furthermore, we discover a novel form fragmentation of dynamics in real space, referred to as \textit{subsystem fragmentation}, seeded by the dissipation which blocks the excitations from propagating across the frozen sites. We analyze these phenomenon analytically and characterize their algebraic properties based on symmetries and their representations. 

Moreover, the solvability of Lindbladian dynamics is typically a hard problem. The generalization of the notions of integrability and fragmentation can allow the discovery of novel dynamical phases of matter. The Hamiltonian is assumed to be frustration-free while the jump operators either commute or anti-commute with each other and the Hamiltonian terms.
We exemplify the structure of symmetries inherited from the Hamiltonian and provide a framework to describe the emergence of new forms of symmetry and fragmentation in the presence of dissipation. We introduce frustration graphs~\cite{Chapman2020,Elman2021,chapman2023,Pozsgay2024} defined in the space of operators to derive effective non-Hermitian Hamiltonians describing the dynamics in each fragment. We utilize the structures of the graphs to classify Lindbladians with free-fermionic solvability, fragmentation, and spectral chaos. This mapping provides a concrete way of engineering and probing non-Hermitian dynamics in operator space in open quantum systems which can potentially be accessed experimentally. Accessing non-Hermitian evolution through operator dynamics is physically more feasible than the no-click limit of quantum trajectories, which requires post-selection of a tiny number of trajectories considered to be exponentially hard.

The rest of the paper is organized as follows. Section~\ref{sec:models} describes the general class of models and their salient properties. In Section~\ref{sec:theory}, we provide a brief overview of Hilbert-space fragmentation in closed quantum systems and its generalization to operator-space fragmentation in open quantum systems. This is followed by a general description of subspace fragmentation in Pauli-Lindblad models and its algebraic properties in Section~\ref{sec:theory}. Subsequently, in Section~\ref{sec:subfrag} we show the formation of subsystem fragmentation in real space, and uncover the effective non-Hermitian dynamics of the fragments using frustration graphs. The spectral and dynamical properties of the fragments and the phase transitions in their non-equilibrium dynamics is described in Section~\ref{sec:conseq}. Finally, in Section~\ref{sec:conclusion} we summarize our results and provide an outlook on future directions.


\section{Dissipative stabilizer models}
\label{sec:models}
The stabilizer formalism is central to a large class of models for quantum error correction (QEC). The toric code is a paradigmatic example where the gapped ground state manifold given by stabilizers is useful for fault tolerance. Stabilizer models are composed of clusters of Pauli operators which can act on qubits in finite dimensional lattices or graphs. The noise can be characterised by dephasing or bit flip operators, which in general can form Pauli channels. This class of models cover a wide range of physical systems being investigated for fault-tolerant quantum computing. In this work we study quantum systems coupled to a Markovian environment described by the Gorini–Kossakowski–Sudarshan–Lindblad master equation~\cite{Gorini1976May,Lindblad1976} (referred to as the Lindbladian in the rest of the paper). 
The evolution of the density matrix representing the mixed state of the system (with $\hbar$ set to $1$ throughout) is given by
\begin{equation}
	\frac{d\rho}{dt}=\mathcal{L}(\rho)= -i\left[H,\rho\right] + \sum_{j} \kappa_j\left(2F_j\rho F_j^\dagger -\{F_j^\dagger F_j, \rho \} \right),
	\label{eq:lindblad}
\end{equation}
and consists of a unitary part governed by a Hamiltonian $H$ and of a dissipative part modelling the coupling to the environment through a set of jump operators $F_j$. 

Open quantum systems are characterised by weak and strong symmetries due to the interplay of unitary and dissipative terms~\cite{PhysRevA.89.022118}. A weak symmetry is a unitary superoperator $\mathcal{U}$ which leaves the Lindbladian invariant: $\mathcal{U}^\dagger\mathcal{L}\mathcal{U}=\mathcal{L}$, hence commuting with the time evolution. On the other hand, a strong symmetry is an operator $J$ such that $\left[ H,J \right]=0=[F_j,J]$  $ \forall j$. The distinction between strong and weak symmetry is crucial for dynamics of mixed states, and we will exemplify their role in fragmentation in Lindbladian models in Sec.~\ref{sec:theory_open}.

In particular, we investigate a class of models with qubits arranged in a one-dimensional lattice of length $N$, with the Hamiltonian and jump operators of the form given below:
\begin{enumerate}
    \item The Hamiltonian is a sum of mutually commuting Pauli strings $h_l$, thus belonging to a stabilizer subgroup (it may also be referred to as a frustration-free Hamiltonian),
    \begin{equation}
        H=\sum_l J_l h_l,\label{eq:Hamilt}
    \end{equation}
    with $\left[h_k,h_l\right]=0$ $\forall k,l$,
    \item The jump operators $F_j$ are chosen from the set of arbitrary Pauli strings, such that $F^\dagger_j=F_j$ and $F^\dagger_j F_j = \mathds{1}$.
\end{enumerate}
A \textit{Pauli string} refers to the tensor product of Pauli operators, denoted $X_l,Y_l$ and $Z_l$ for site $l$, over the lattice of spins. If the jump operators commute with the Hamiltonian terms, the system is trivially solvable. Therefore, in order to investigate complex dynamics in these models, we assume them to be non-commuting.

There is a subset of size $M\leq N$, of the Hamiltonian terms that generate the rest of the terms, which are the the stabilizer generators. They generate an Abelian subgroup, which can be mapped (non-uniquely) by means of a Clifford transformation to a basis in which each generator is represented by $\tz_l$. We can also define a set of destabilizer operators $\tx_l$ which pairwise anticommute with each of the stabilizer generators. The destabilizers can be recovered in the original basis by performing the inverse transformation. In this new basis, any Hamiltonian of interest can be written as a sum over products of generators:
\begin{equation}
    H=\sum_{\Vec{n}} J_{\Vec{n}}  \prod_{l=1}^M \tz^{n_l}_l,
    \label{eq:Hstabgen}
\end{equation}
where the sum is over a given set of vectors $\Vec{n}=(n_1,n_2,...,n_M)$, with $n_l=0,1$.  

This basis also contains $N-M$ \textit{free} stabilizers $\tz_l$ for $l>M$ which do not appear in the Hamiltonian. These are operators that, together with the generators of~\eqref{eq:Hstabgen}, form a complete basis for the operator-space. Also, they commute or anticommute with all the dissipative terms in the Lindblad equation, since the jumps are Pauli strings in this new basis. The same applies to the corresponding conjugate $\tx_l$'s. The operators anticommuting with the jump operators form a set of weak symmetries acting on $\rho$ as $\tz_l \rho \tz_l$, whereas the commuting ones form a set of strong symmetries. Despite the simplicity of this class of models, they can host a wide range of universal dynamical behaviour. 

In order to illustrate the theory and explore its consequences, we consider a toy model with a $ZXZ$ Hamiltonian on a one-dimensional qubit chain of length $N$~\cite{PhysRevLett.59.799,PhysRevB.45.304,Kennedy1992}:
\begin{equation}
	H=\sum_{l=2}^{N-1} J_l Z_{l-1}X_l Z_{l+1},
	\label{eq:Hzxz}
\end{equation}
where $X_l$ and $Z_l$ denote Pauli operators at site $l$. 
In this model, the generators are $\tz_l=Z_{l-1}X_l Z_{l+1}$ for $2\leq l\leq N-1$, and there are two free degrees of freedom at the edges: $\tz_1 = X_1Z_2$ and $\tz_N = Z_{N-1}X_N$.
This model possesses a symmetry-protected topological phase~\cite{Bahri2015,PhysRevB.104.014424,PhysRevLett.125.200506,Son_2011}, due to spin flips at odd and even sites that generate a global $\mathds{Z}_2\cross\mathds{Z}_2$ symmetry.  
Recently, we have studied its topological properties under coupling to an environment~\cite{PRXQuantum.5.030304}.

\begin{figure*}[!ht] 
	\centering
    \includegraphics{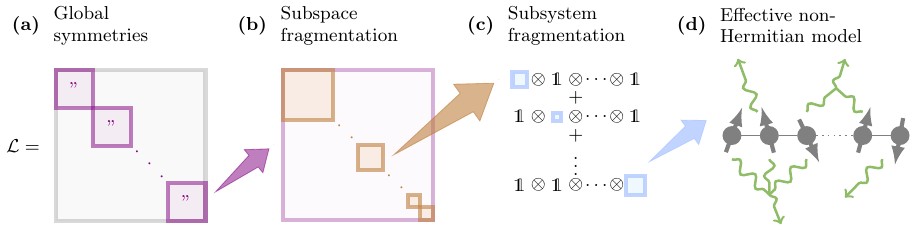}
	\caption{textit{Generation of subspace and subsystem fragmentation:} Schematic picture of hierarchical fragmentation of the spectrum of Pauli-Lindblad models. (a) Global symmetries partition the state-space into sectors such that the Lindbladian becomes block-diagonal. (b) The algebraic structure of the models causes the decomposition of each sector into independent subspaces, giving rise to \textit{subspace fragmentation}. (c) In each fragment, the presence of \textit{blockades} leads to \textit{subsystem fragmentation} in real space, where the degrees of freedom in each subsystem evolve independently. (d) In the fragmented state-space, the action of the Lindbladian on density matrices can be mapped to an effective non-Hermitian model acting on pseudospins. This hierarchical decomposition simplifies the complexity of dynamics of open quantum systems.}
	\label{fig:steps}
\end{figure*}

In what follows, we will show that the operator space, in which the density matrices representing mixed states exists, exhibits two classes of fragmentation with a hierarchical structure. Firstly, the operator space splits into a direct sum of fragments, which we refer to as  \textit{subspace fragmentation}. The algebraic properties of the Hamiltonian and jump operators control the nature of the fragments in operator space and the dynamics within them. These fragments can further split into subsystems in real space, where the dynamics between different subsystems are decoupled, a phenomenon referred to as \textit{subsystem fragmentation}. The degrees of freedom in real space separate between frozen sites with local conservation laws, and active sites which remain dynamical. In each of these real-space fragments, the dynamics is governed by an effective non-Hermitian Hamiltonian, which exhibits range of properties from integrability to quantum chaos. A schematic picture of the phenomenology is illustrated in Fig.~\ref{fig:steps}.

We will use jump operators in the form of Pauli-strings given by 
\begin{equation}
    F_j=\bigotimes_{i=1}^L X_i^{n^j_i} Z_i^{n^j_{L+i}}, \label{eq:Fjumps}
\end{equation}
where $\Vec{n}^j$ is a binary vector of length $2L$.
In particular, we use the following examples to exemplify the organizing principles of fragmentation and their dynamics in open quantum systems:
\begin{itemize}
    \item \textit{Integrable fragmentation:} The jump operators, $F_j=Z_{j-1}Z_{j+1}$ for $j=2,...,N-1$, lead to subspace and subsystem fragmentation where the dynamics in each fragment can be mapped to a free-fermionic non-Hermitian Hamiltonian. In this case, the  $\mathds{Z}_2\cross\mathds{Z}_2$ symmetry of the cluster model is strong, and there are 2 additional strong symmetries acting as edge mode operators; $Z_1$ and $Z_N$. 
    \item \textit{Chaotic fragmentation:} The jump operators, $F_j=Y_j$ for $j=1,...,N$, lead to chaotic subspace and subsystem fragmentation. In this case, the dissipation breaks the global $\mathds{Z}_2\cross\mathds{Z}_2$ symmetry of the Hamiltonian to a weak symmetry.
\end{itemize}
Some of the authors studied the cluster model with above-mentioned jump-operators in an earlier work~\cite{PRXQuantum.5.030304} which showed the stability of symmetry protected topological order and the associated dynamics of edge states in the presence of dissipation. The topological dynamics of density matrices were characterized by the low lying spectrum of the Lindbladian combined with the evolution of pure-state quantum trajectories. In this work, our choice of the form of jump operators covers a range of dynamical regimes from integrability to chaos. Henceforth, we develop a general framework which describes a large class of Lindbladian models and their spectral properties. 

\section{Theory of operator-space fragmentation}
\label{sec:theory}

The general phenomenon of subspace fragmentation is rooted in the algebraic structure of the open quantum system. In isolated systems, the framework of Hilbert space fragmentation builds on the bond and commutant algebras originating from the terms in the Hamiltonian. In order to generalize this framework in the presence of dissipation, we develop an analogous description in the space of superoperators and their algebraic properties. The superoperator algebra is naturally suited for treating strong and weak symmetries of the Lindbladian on an equal footing and classifying its spectral properties.   
Here, we briefly present the framework in closed systems and continue to elucidate its generalization to open systems. 
In Section~\ref{sec:worked_examples}, we will apply this theory to show subspace fragmentation acts in the case of general Pauli-Lindblad models and characterize their spectral and dynamical properties in the specific instances of the dissipative cluster model.

\subsection{Closed systems}

Hilbert-space fragmentation in closed quantum systems~\cite{PhysRevX.10.011047,PhysRevB.101.174204,Moudgalya_2022} has a rigorous definition using the theory of bond and commutant algebras~\cite{PhysRevX.12.011050}. 
The bond and commutant algebras are defined as,
\begin{align}
    \mathcal{A}&=\langle \{h_l\} \rangle, \\
    \mathcal{C}&=\{ O \; | \; [O, h_l]=0 \; \forall l  \},
\end{align}
where $h_l$ is a term in the Hamiltonian. The bond algebra $\mathcal{A}$ is defined as the algebra generated by arbitrary linear combinations of arbitrary products of the Hamiltonian terms $h_l$ and the identity, which is denoted by $\langle\dots\rangle$.
All operators that commute with each of these local Hamiltonian terms form a closed associative algebra called the commutant algebra $\mathcal{C}$, which can confer a fragmented block-diagonal structure to $H$, following Schur's lemma.
The von Neumann bicommutant theorem allows the Hilbert space to be decomposed in a manner similar to the Schur-Weyl duality,
\begin{equation}
    \mathcal{H}=\bigoplus_\lambda \left( \mathcal{H}^\mathcal{C}_\lambda \otimes \mathcal{H}^\mathcal{A}_\lambda \right),
\end{equation}
where $\mathcal{H}_{\lambda}^{\mathcal{A}/\mathcal{C}}$ are distinct irreducible representations of either $\mathcal{A}$ or $\mathcal{C}$, both labeled by $\lambda$, of dimensions $D_{\lambda}$ and $d_{\lambda}$, respectively.
If we then choose some orthonormal basis for each of $\mathcal{H}_\lambda^\mathcal{C}$ then each basis vector identifies a $H$-invariant subspace called a Krylov subspace.

The Hilbert space can decompose into sectors due to the presence of symmetry or non-trivial topology, but these are not considered to be a signature of fragmentation.
In these cases, the commutant algebra can contain the associated symmetry generators or logical operators in topologically ordered subspaces, but these produce at most a polynomial number of Krylov subspaces.
The system is said to be fragmented when the number of disconnected subspaces scales exponentially with system size, which is generated by the remaining elements of the commutant.
The structure of the commutant algebra constraints the size and distribtuion of the subspace: $\dim \mathcal{C}$ is exponential in system size if and only if the number of fragments is exponential.


\subsection{Open systems}\label{sec:theory_open}

Now, let us generalize the theory of bond and commutant algebras to open quantum systems, in which the fragmentation occurs in the space of operators. The general Lindbladian evolution can be expressed as a sum of commutators $u_l(\rho)=\left[ h_l,\rho \right]$ due to unitary dynamics and dissipative terms due to individual jump operators $d_j(\rho)=2F_j\rho F_j^\dagger -\{F_j^\dagger F_j, \rho \}$.
In this form, \eqref{eq:lindblad} becomes 
\begin{equation}
	\mathcal{L}(\rho)= -i\sum_l J_l u_l(\rho) + \sum_{j} \kappa_j d_j(\rho).
	\label{eq:lindblad2}
\end{equation}
The parameters $J_l$ and $\kappa_j$ can be arbitrary.

Open systems are characterized by weak and strong symmetries, as defined in Sec.~\ref{sec:models}. In earlier work, fragmentation was considered by promoting the symmetries of the Hamiltonian to strong symmetries by choosing the jump operators without breaking those symmetries~\cite{PhysRevResearch.5.043239}, which retains the algebraic structure of the commutants in the isolated system. 
In this work, we develop a theory of fragmentation for both strong and weak symmetries by working with algebras of superoperators, and their consequences for the spectrum and spatio-temporal structure of the mixed states.



We define the bond algebra as the algebra that contains arbitrary linear combinations of arbitrary products of superoperator terms of the Lindbladian~\eqref{eq:lindblad2}, and the commutant as all superoperators that commute with all the bond elements,
\begin{align}
    \mathcal{A}&=\langle \{u_l\}, \{d_j\}  \rangle,\\
    \mathcal{C}&=\{ O \; | \; [O, u_l]=[O, d_j]=0 \; \forall l,j  \}.\label{eq:comm_def}
\end{align}
As in the closed system, the Hilbert-Schmidt space of bounded operators $\mathcal{B}(\mathcal{H})$ can be decomposed using irreducible representations of the two algebras defined above,
\begin{equation}
    \mathcal{B}=\bigoplus_\lambda \left( \mathcal{B}^\mathcal{C}_\lambda \otimes \mathcal{B}^\mathcal{A}_\lambda \right).
\end{equation}
This means that the commutant elements can be written as
\begin{align}
O = \bigoplus_\lambda \left(O_\lambda \otimes \mathds{1}_\lambda^\mathcal{A}\right). \label{eq:Ocomm}
\end{align}
The Lindbladian $\mathcal{L}$, being in the bond algebra, needs to commute with the operator $O$, so it acquires a corresponding block diagonal structure as shown in \figref{fig:steps}[b] by Schur's lemma,
\begin{align}
    \mathcal{L} &= \bigoplus_\lambda \left(\mathds{1}_\lambda^\mathcal{C} \otimes L_\lambda\right). \label{eq:Lblocks}
\end{align}

Operator-space fragmentation has direct consequences for the dynamics of the system. 
Just as global symmetries prevent from mixing operators that transform differently under those symmetries, fragmentation prevents operators from escaping the fragment the system belongs to during the time evolution. While the algebraic framework describes the subspace fragmentation, the open quantum system can also exhibit \textit{subsystem fragmentation}--where fragments further decouple into spatially separated regions (\figref{fig:steps}[c])--arising from additional dynamical constraints not captured by the commutant algebra. The localized dissipation effectively generates local conserved quantities which act as \textit{blockades} and partition fragments into dynamically independent subsystems in real space. In order to develop a complete theory of operator-space fragmentation in open systems requires supplementing the algebraic theory with graph-theoretic methods to identify the underlying spatial structure of the mixed states.  In the following sections, we apply the framework of frustration-graphs (Sec.~\ref{sec:frustgraphs}) to exemplify this phenomena in the class of Pauli-Lindblad models as defined in Sec.~\ref{sec:models}.

\section{Subspace Fragmentation}
\label{sec:worked_examples}


In this section, we apply the superoperator algebra formalism to the Pauli-Lindblad models defined by the Hamiltonian terms of the form in Eq.~\eqref{eq:Hstabgen} and dissipative terms given by~\eqref{eq:Fjumps}. The fragmentation of the operator space is a general property of this family of models. We demonstrate it by characterizing the commutant algebra and its irreducible representations, which we refer to as irreps in the rest of the paper.

In the following, we work in the basis in which the Hamiltonian takes the form of~\eqref{eq:Hstabgen}. That is, the basis of the operator space is determined by the stabilizer generators $\tz_l$ present in the Hamiltonian. Each generator has an associated Pauli space spanned by $\{\tx_l,\ty_l,\tz_l,\mathds{1}_l\}$, and the total space of $N$ spins is the tensor product of all of them. If there are $M<N$ generators, there is freedom to choose the remaining $N-M$ degrees of freedom to complete the basis of Pauli strings.

The bond algebra contains, on the one hand, unitary evolution terms $u_l$, which are commutators with the stabilizer generators and possibly with arbitrary products of stabilizer generators. It also  contains Pauli-string dissipative terms $d_l$, which act diagonally in the chosen basis. 
In this section, the symmetries of these models are studied through the commutant algebra of superoperators, and we proceed as follows. In Sec.~\ref{sec:comm_struct}, we describe the commutant and its maximal Abelian subalgebra. In Sec.~\ref{sec:comm_irreps}, we show how to find the irreps $\lambda$ of the commutant. And finally, in Sec.~\ref{sec:comm_frag}, we characterize the block-diagonal form of the Lindbladian over the fragmented space, as shown in~\eqref{eq:Lblocks}.

\subsection{Structure of the commutant}\label{sec:comm_struct}

Here, we consider the superoperators that can be included in the commutant, given all the elements of the bond algebra. We distinguish the constraints on these superoperators originating from the stabilizer generators of the Hamiltonian and the free degrees of freedom.

\subsubsection{Stabilizer generators}

The unitary time-evolution is generated by the stabilizer generators. Each generator $\tz_l$ of the Hamiltonian~\eqref{eq:Hstabgen} has an associated commutant algebra. It can be understood by observing that the following superoperator projectors onto $\ti_l$ and $\tz_l$ are conserved by the dynamics--they commute with the bond algebra. They have a local form and are given by
\begin{equation}
    \begin{alignedat}{2}
        &P_{\ti_l} (\cdot) &&= \frac{1}{4} \left(\mathds{1}+\tx_l \cdot \tx_l +\ty_l \cdot \ty_l +\tz_l \cdot \tz_l\right), \\
        &P_{\tz_l} (\cdot) &&= \frac{1}{4} \left(\mathds{1}-\tx_l \cdot \tx_l -\ty_l \cdot \ty_l +\tz_l \cdot \tz_l\right).
    \end{alignedat}\label{eq:projs}
\end{equation}
The two conserved superoperators \eqref{eq:projs} generate an algebra for each generator $l$, by all possible products and sums of projectors (plus the identity). A basis for this algebra, $\mathcal{C}^\mathrm{g}_l \; (l\leq M)$, is simply $\{\mathds{1}, P_{\ti_l}, P_{\tz_l} \}$.
The algebras associated with different generators are independent, such that the total algebra due to all of them is $\mathcal{C}^{\mathrm{g}}=\bigotimes_{l=1}^{M} \mathcal{C}^\mathrm{g}_l$.

In the following, we will use the irreducible representations of these algebras. 
In order to find the irreps of $\mathcal{C}^\mathrm{g}_l$, we find a basis of the commutant that forms a group, identifying it as isomorphic to the group algebra $\mathbb{C}[\mathbb{Z}_3]$.
The associated group $\mathbb{Z}_3$ has the same irreps, due to the isomorphism.
This correspondence is established by finding a cyclic element $R_l$ which generates the entire algebra:
\begin{equation}
    R_l=\mathds{1}-(1-\omega)P_{\ti_l} - (1-\omega^2)P_{\tz_l},\label{eq:R_l}
\end{equation}
where $\omega=e^{i2\pi/3}$.
It can easily be checked that $\{ \mathds{1}, R_l, R^2_l \}$ forms a basis for the algebra of projectors.

\subsubsection{Free degrees of freedom}
There are $N-M$ free degrees of freedom, each with an associated algebra $\mathcal{C}^\mathrm{f}_l \; (l>M)$. As these degrees of freedom are free, $\mathcal{C}^\mathrm{f}_l$ contains all possible commuting superoperators, for which the following diagonal basis can be chosen, 
\begin{equation}
\{\mathds{1}, \tx_l \cdot \tx_l, \ty_l \cdot \ty_l, \tz_l \cdot \tz_l \}. \label{eq:Z2xZ2basis}
\end{equation}
They are also all independent, such that $\mathcal{C}^{\mathrm{f}}=\bigotimes_{l=M+1}^{N} \mathcal{C}^\mathrm{f}_l$. 

The basis~\eqref{eq:Z2xZ2basis} forms a $\mathbb{Z}_2\times\mathbb{Z}_2$ group, which can be seen by taking $\tx_l\cdot \tx_l$ and $\ty_l\cdot \ty_l$ as the generators of the two $\mathbb{Z}_2$ groups. Therefore, each $\mathcal{C}^\mathrm{f}_l$ is isomorphic to the group algebra $\mathbb{C}[\mathbb{Z}_2\times\mathbb{Z}_2]$.

\subsubsection{Maximal Abelian subalgebra of $\mathcal{C}$}

Finally, the maximal Abelian subalgebra of the commutant is $\mathcal{C}^{\text{max}}=\mathcal{C}^{\mathrm{g}}\otimes\mathcal{C}^{\mathrm{f}}$. 
It is sufficient to study the maximal Abelian subalgebra, because the Lindbladian is put into block-diagonal form over the fragmented space by choosing a set of mutually commuting symmetry superoperators, whose eigenvalues label the blocks. That maximal set of symmetries is the maximal Abelian subalgebra $\mathcal{C}^{\text{max}}$, and could be non-unique. We note that $\mathcal{C}$ can be non-Abelian, and thus bigger than $\mathcal{C}^{\text{max}}$, which would add structure beyond fragmentation: it can lead to degeneracies in the spectrum of $\mathcal{L}$ because of the presence of irreps of dimensions bigger than $1$.

Now, we prove that $\mathcal{C}^{\mathrm{g}}\otimes\mathcal{C}^{\mathrm{f}}$ is a maximal Abelian subalgebra of $\mathcal{C}$. First, we notice that the action of all of its elements is diagonal, therefore it is Abelian. Second, we show that it is maximal, that is, that it cannot be enlarged without losing the Abelian property. For this, it is useful to change the basis associated with stabilizer generators in order to make the corresponding $u_l(\cdot) = [\tz_l,\cdot]$ in the bond algebra diagonal: $\{\mathds{1},\tz,\tx,\ty\}_l \longrightarrow \{(i\tx+\ty)/2,(-i\tx+\ty)/2,\tz,\mathds{1}\}_l$. In this basis, $\mathcal{C}^\mathrm{g}_l$ and $\{u_l\}$ span all possible diagonal superoperators. $u_l$ itself cannot be added to the commutant, because in general, it does not commute with the dissipators $d_j$ in the bond algebra (assumption we make to keep the model non-trivial). As to $\mathcal{C}^\mathrm{f}_l$ algebras, they form a complete set of diagonal matrices, maximally Abelian by definition. Therefore, $\mathcal{C}^{\text{max}}$, tensor product of $\mathcal{C}^\mathrm{g}_l$'s and $\mathcal{C}^\mathrm{f}_l$'s, already contains all diagonal superoperators, except for those spanned by $\{u_l\}$, which cannot be added. The only way to expand $\mathcal{C}^{\text{max}}$ would be by adding a non-diagonal superoperator. However, a non-diagonal superoperator cannot simultaneously commute with all possible diagonal superoperators, which $\mathcal{C}^{\text{max}}$ and $\{u_l\}$ contain. The added element is guaranteed either to break the Abelian property because it does not commute with $\mathcal{C}^{\text{max}}$, or not to be a valid element of $\mathcal{C}$ because it does not commute with $\{u_l\}$ from the bond algebra. This proves maximality.

\subsection{Irreducible representations of the commutant}\label{sec:comm_irreps}

Now, we show that the structure of the commutant described above leads to fragmentation of the space of operators. This is accomplished by finding the irreps of $\mathcal{C}^{\text{max}}$. These are tensor products of the irreps of $\mathcal{C}_l^\mathrm{g}$ and $\mathcal{C}_l^\mathrm{f}$.

First, we work with the generator algebras $\mathcal{C}_l^\mathrm{g}$. We want to reduce $\rho_l$, the Pauli space associated to generator $l$, into irreps of the $\mathbb{Z}_3$ group defined by the cyclic element~\eqref{eq:R_l}. This can be done by projecting $\rho_l$ onto the irreps using the character table in Table~\ref{tab:chi}. The character of a group representation $\rho$ is a function on the group $G$ that associates to each group element $g\in G$ the trace of the corresponding representation matrix: $\chi_\rho(g) = \operatorname{Tr}(\rho(g))$.

\begin{table}[h]
\caption{\label{tab:chi}
Character table of the 3 irreducible representations of $\mathbb{Z}_3$ as well as of the representation $\rho_l$ (the local Pauli-space). Columns are the different conjugacy classes.}\begin{ruledtabular}
\begin{tabular}{l|ccc} 
& $\{\mathds{1}\}$ & $\{R_l\}$ & $\{R_l^2\}$ \\
\colrule
$\chi_{\omega^0}$ & $1$ & $1$ & $1$ \\
$\chi_{\omega^1}$ & $1$ & $\omega$ & $\omega^2$ \\
$\chi_{\omega^2}$ & $1$ & $\omega^2$ & $\omega$ \\ 
$\chi_{l}$ & $4$ & $2+\omega+\omega^2=1$ & $2+\omega+\omega^2=1$ \\
\end{tabular}
\end{ruledtabular}
\end{table}

The inner-product between characters is defined as
\begin{equation}
    \langle \chi_i, \chi_j\rangle = \frac{1}{|G|}\sum_{g\in G} \chi_i(g) \chi^*_j(g),
\end{equation}
where $|G|$ is the order of the group $G$. In our case, using Table~\ref{tab:chi}, this yields $\langle \chi_l, \chi_{\omega^0}\rangle = 2$ and $\langle \chi_l, \chi_{\omega^1}\rangle = \langle \chi_l, \chi_{\omega^2}\rangle = 1$, which means that the trivial irrep appears twice in the decomposition, while the two non-trivial irreps appear once each.

Therefore, under this $\mathbb{Z}_3$ action, the local Pauli space at site $l$ decomposes into irreps as 
\begin{equation}
    \rho_{l} \simeq  (\mathbb{C}^2 \otimes \rho_{\omega^0}) \oplus \rho_{\omega^1} \oplus \rho_{\omega^2}, \label{eq:rho_l}
\end{equation} 
where the different one-dimensional representations are labeled by the powers of $\omega$. The $\mathbb{C}^2$ factor reflects the twofold multiplicity of the trivial irrep ($\rho_{\omega^0}$). It corresponds to the ${\tx,\ty}$ subspace where the symmetry algebra acts trivially.

Second, we proceed analogously for the irreps of $\mathcal{C}^\mathrm{f}_l$. The basis \eqref{eq:Z2xZ2basis} forms the $\mathds{Z}_2\times \mathds{Z}_2$ group. The four-dimensional representation on that local Pauli space decomposes into a direct sum of the four one-dimensional irreps of that group. This can be seen by reasoning with the $\mathds{Z}_2\times \mathds{Z}_2$ character table, analogously to what was done for $\mathds{Z}_3$ above.

To sum up, the algebra $\mathcal{C}^{\text{max}} = (\bigotimes_{l=1}^M \mathcal{C}^\mathrm{g}_l) \otimes (\bigotimes_{l=M+1}^N \mathcal{C}^\mathrm{f}_l)$ is a tensor product of local subalgebras, where $\mathcal{C}^\mathrm{g}_l$ (for $l \leq M$) and $\mathcal{C}^\mathrm{f}_l$ (for $l > M$) correspond to stabilizer generators and free degrees of freedom, respectively. Its irreps are tensor products of irreps of $\mathcal{C}^\mathrm{g}_l$ and $\mathcal{C}^\mathrm{f}_l$. 
Each irrep $\lambda$ of $\mathcal{C}^{\text{max}}$ is labeled by a tuple 
\begin{equation}
    \lambda = (\lambda_1, \ldots, \lambda_M, \lambda_{M+1}, \ldots, \lambda_N), \label{eq:lambda_tuple}
\end{equation} 
where $\lambda_l$ denotes the irrep of $\mathcal{C}^\mathrm{g}_l$ (for $l \leq M$) or $\mathcal{C}^\mathrm{f}_l$ (for $l > M$). For $l \leq M$, $\mathcal{C}^\mathrm{g}_l$ has three 1D irreps. For $l > M$, $\mathcal{C}^\mathrm{f}_l$ has four 1D irreps. The dimension of the irrep $\lambda$ is $D_\lambda = \prod_{l=1}^M d_{\lambda_l} \prod_{l=M+1}^N d'_{\lambda_l}$, where $d_{\lambda_l}$ and $d'_{\lambda_l}$ are the dimensions of the local irreps $\lambda_l$. In our case, all the irreps are 1D, so that $D_{\lambda}=1$ for all $\lambda$. 

\subsection{Structure of the corresponding fragments}\label{sec:comm_frag}

The total space, tensor product of all local Pauli spaces, is decomposed into irreps of $\mathcal{C}^{\text{max}}$. The irreps that involve a tensor product of $k$ local trivial irreps of $\mathcal{C}^\mathrm{g}_l$ (i.e of $\mathbb{Z}_3$) appear with a multiplicity $m_\lambda$ of $2^k$ in this decomposition. This follows from the twofold multiplicity contributed by a trivial irrep on each site, as in \eqref{eq:rho_l}.

Any operator $O \in \mathcal{C}^{\text{max}}$ decomposes as in Eq.~\eqref{eq:Ocomm}, where $O_\lambda$ are matrices of size $D_\lambda \times D_\lambda$, and $\mathds{1}^{\mathcal{A}}_{\lambda}$ are identity matrices of size $m_\lambda \times m_\lambda$, corresponding to the multiplicity of the given representation.

$\mathcal{C}$ does not distinguish between the $m_{\lambda}$ copies of irrep $\lambda$, allowing the bond algebra to mix them freely. That ensures that the Lindbladian $\mathcal{L}$ inherits the block structure of Eq.~\eqref{eq:Lblocks}, where each $L_\lambda$ acts on the $m_\lambda$-dimensional space.

The fragments correspond to irreps $\lambda$. The size of the $L_\lambda$ blocks is the corresponding multiplicity $m_\lambda = 2^k$, where $k=0,1,...,M$ counts the number of sites $l$ on which $\mathcal{C}^\mathrm{g}_l$ ($l \leq M$) acts trivially. 
The number of fragments with size $2^k$ is computed as follows: we choose $k$ of the $\mathcal{C}^\mathrm{g}_l$ terms contributing a factor of 2 to the multiplicity, the remaining $M-k$ terms have two possibilities, and the $N-M$ $\mathcal{C}^\mathrm{f}_l$ terms have 4 possibilities:
\begin{equation}
    \binom{M}{k} 2^{M-k} 4^{N-M}. \label{eq:frag_num_1}
\end{equation}
Summing over $k$ gives the total number of fragments,
\begin{equation}
    \sum_{k=0}^M \binom{M}{k} 2^{M-k} 4^{N-M} = 3^M 4^{N-M}, \label{eq:frag_num_tot}
\end{equation}
matching the dimension of $\mathcal{C}^{\text{max}}$. The number of independent subspaces scales exponentially with $N$, confirming fragmentation.

Finally, we can verify that this decomposition yields a complete set of fragments covering the full $4^N$-dimensional space, by summing over the sizes of all the fragments,
\begin{equation}
    \sum_{k=0}^M \binom{M}{k} 2^{M-k} 4^{N-M} \times 2^k = 4^{N}. \label{eq:space_tot}
\end{equation}

\subsection{fragmentation in dissipative $ZXZ$ cluster model}
\label{sec:cluster}

Here we elucidate the formation of subspace fragments in the cluster model~\eqref{eq:Hzxz} with Pauli string jump operators and link it to the above theory. In this class of models, the fragmentation of the operator space, represented in \figref{fig:steps}[b], occurs in the basis of Pauli strings. It stems from the frustration-free form of the Hamiltonian where all terms commute with one another, combined with the action of the jump operators on elements of the Pauli string basis. We introduce a change of basis that transforms the three-body interactions to a single qubit Pauli-$Z$.
In the bulk, for $l=2,...,N-1$, the stabilizer basis is transformed as follows
\begin{equation}
    \begin{alignedat}{2}
        &\tz_l &&=Z_{l-1}X_l Z_{l+1},\\
        &\tx_l &&= Z_l. 
    \end{alignedat} \label{eq:tildebasis1}
\end{equation}
At the boundaries for $l=1$ and $N$, we similarly define
\begin{equation}
    \begin{alignedat}{4}
        &\tz_1 &&=X_1 Z_{2}, &&\quad \quad  \tz_N &&=Z_{N-1} X_{N}, \\
        &\tx_1 &&= Z_1,&&\quad \quad \tx_N &&= Z_N. 
    \end{alignedat} \label{eq:tildebasis2}
\end{equation}
The corresponding $\ty_l$ operators are given by $\tz_l\tx_l=i\ty_l$, such that $\{\tx_l,\ty_l, \tz_l\}$ form an su(2) algebra at each site $l$. Note that any Pauli string in the original basis will remain a Pauli string in this new basis.

Now, let us look at the action of each term of the Lindbladian on a basis element $\tm$ -- a Pauli string of length $N$. 
A unitary term given by $u_l(\tm)=[\tz_l,\tm]$ acts locally on site $l$ of $\tm$, denoted by $\tm_l$, and does not affect other sites,
\begin{equation}
    u_l(\tm_l)=\left\{\begin{alignedat}{2}
        &0 \quad &&\text{if } \tm_l=\ti_l \text{ or }\tz_l,\\
        &2i\ty_l \quad &&\text{if } \tm_l=\tx_l,\\
        &-2i\tx_l \quad &&\text{if } \tm_l=\ty_l.
    \end{alignedat}\right. \label{eq:unitary_action}
\end{equation}
The action of the dissipative terms which we consider to be Pauli strings is given by $d_j(\tm)=2( \tilde{F}_j \tm \tilde{F}_j - \tm)$. Conjugation of a Pauli string by another Pauli string can only change the sign, $\tilde{F}_j \tm \tilde{F}_j=\varepsilon_j \tm$ with $\varepsilon_j=\pm 1$, hence
\begin{equation}
    d_j(\tm)=\left\{\begin{alignedat}{2}
        &0 \quad &&\text{if } \varepsilon=1,\\
        &-4\tm \quad &&\text{if } \varepsilon=-1,
    \end{alignedat}\right. \label{eq:dissip_action}
\end{equation}
-- the dissipation acts diagonally in this basis and plays no role in fragmentation. 


Therefore, the action of $\mathcal{L}$ on a Pauli string basis element $\tm$ does not affect the sites with states $\ti$ or $\tz$, which we refer to as \textit{frozen} sites because they are conserved by the dynamics. This fact was formally expressed through the conserved superoperator projectors in \eqref{eq:projs}. On the other hand, this operation can also create a superposition of Pauli strings where each term differs from $\tm$ by a single site flip: $\tx_l \leftrightarrow \ty_l$. These dynamical degrees of freedom form a collection of \textit{active} sites and are not distinguishable by the symmetry algebra, on which it acts trivially (see Eq.~\eqref{eq:rho_l}). These degrees of freedom are susceptible to the dynamics, as they are not fixed by the symmetries.

The repeated action of $\mathcal{L}$ on $\tm$ generates an invariant Krylov subspace. It contains all the operators with an identical pattern of frozen sites, and all possible configurations of active sites, each one being in either $\tx$ or $\ty$. Note that since the Hamiltonian~\eqref{eq:Hzxz} does not act on the boundaries $l=1$ and $l=N$, these sites remain frozen independent of their state. 
An example for a chain of length $N=8$ and a single frozen site in the bulk and two on the edges is
$$\tm = {\color{gray}\tz} \; {\color{myred}\tx \; \ty}\; {\color{gray}\ti}\; {\color{myred}\tx \;\ty\; \tx}\; {\color{gray}\ty},$$
which generates a Krylov subspace of size $2^5=32$ with $5$ active sites (in red).

The number of Krylov subspaces and their dimensions is obtained as follows. Each subspace is uniquely labeled by its pattern of frozen sites, whose number in the bulk is denoted by $k$ and ranges from $0$ to $N-2$, as the boundary sites are frozen by construction. Therefore, the number of fragments with $k$ frozen sites from $N-2$ bulk sites is $ \binom{N-2}{k} 2^k 4^2 $.
The total number of Krylov subspaces is obtained by summing over all possible values of $k$:
\begin{equation}
    \sum_{k=0}^{N-2} \binom{N-2}{k} 2^{k+4} = 4^2 \times 3^{N-2}.
    \label{eq:frag_num_tot_cluster}
\end{equation}
This matches the general expression \eqref{eq:frag_num_tot} with $M=N-2$ and grows exponentially with system size in agreement, a signature of strong fragmentation~\cite{PhysRevX.12.011050}.


\section{Subsystem fragments and effective non-Hermitian models}
\label{sec:subfrag}
Each Krylov subspace of the fragmented space can now be characterized by an effective model describing the dynamics. In this section, we will show how this can be achieved systematically using frustration graphs~\cite{Chapman2020,Elman2021,chapman2023,Pozsgay2024}. 
We will first describe the general concept of frustration graphs and its application for closed quantum systems, in particular their relationship to free-fermionic solutions of many-body problems. Then, we will show how it applies to open systems. And finally, we will generalize the formalism of frustration graph for open quantum systems, and systematically construct effective non-Hermitian Hamiltonians corresponding to the models from Section~\ref{sec:models} with particular choices of dissipation channels. To be specific, each of the subspace fragments, as identified in Section~\ref{sec:worked_examples}, can be characterized by such a Hamiltonian. 
In this way, we will see that different choices of jump operators can lead to drastically different physical behaviour.

\subsection{Frustration graphs}
\label{sec:frustgraphs}

Quantum many-body spin problems are generally difficult to solve exactly. Numerical solutions are challenging as the computational complexity scales exponentially with system size. A certain subset of these problems can be mapped onto non-interacting fermions, for example using the Jordan-Wigner transformation~\cite{Jordan1928}, which reduces the problem to finding the eigenvalues of a matrix that now scales linearly with system size. However, it has been noticed that certain models for which the Jordan-Wigner transformation does not work in the aforementioned manner still have spectral properties analogous to free fermions~\cite{Fendley2019,Elman2021}. They have been called free fermions \textit{in disguise}.

\begin{figure}[!ht] 
	\centering
    \includegraphics{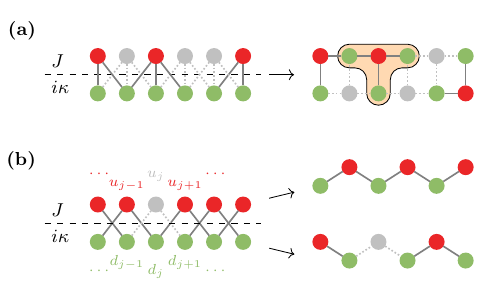}
	\caption{Examples of frustration graphs for the cluster model with (a) $Y$ and (b) $ZIZ$ jump operators. The gray vertices correspond to frozen degrees of freedom in a given fragment, and the red and green vertices represent the unitary and dissipative terms of the Lindbladian respectively. (a) By twisting every other rung of this frustration graph, it takes the form of a ladder. In this case, at least two adjacent vertices need to be frozen in order to form a blockade. It also may contain forbidden subgraphs, such as the claw highlighted in orange, which would mean that it cannot be mapped onto a free-fermion model. (b) The graph decomposes into odd and even chains, and each of them has the structure of the Ising model, solvable by mapping to free fermions. Moreover, one frozen vertex acts as a blockade and decomposes the chain into two independent subsystems.}
	\label{fig:frustration}
\end{figure}

Frustration graphs provide a theoretical framework to characterize non-trivial mappings to free fermions, if they exist, or assert that such a mapping is impossible. 
The Hamiltonian of a spin system can in general be written as a sum of Pauli-string terms, such that all terms either commute or anticommute with each other.
To build a frustration graph, a vertex is assigned to each term in the Hamiltonian and edges are drawn between pairs of vertices corresponding to anticommuting terms.
The given spin system corresponds to a model of non-interacting fermions if its frustration graph satisfies certain properties~\cite{Chapman2020,chapman2023}. 
For example, it cannot contain any claws as subgraphs. 
A claw is a graph comprising a central vertex connected to three other vertices, such as highlighted in orange in \figref{fig:frustration}[a]. 
In practical terms, the presence of certain forbidden subgraphs makes it impossible to find a free-fermion description of the problem through a Jordan-Wigner-like transformation~\cite{Chapman2020}. However, this is not the complete picture, as there also exist more complex mappings of a spin system onto free fermions. A more general framework has been developed in Ref.~\cite{chapman2023}.

In open quantum systems, the superoperators $u_l$ and $d_j$ of the Lindblad master equation~\eqref{eq:lindblad2} can also be written as linear combinations of Pauli strings acting in the space of operators. We show that these superoperators can be used to construct a frustration graph in close analogy with isolated systems.

Constructing the graph is straightforward due to the simple form of these operators.
Namely, jump operators are Pauli strings, which implies that the second term of the dissipators $d_j(\rho)=2F_j \rho F_j - 2\rho$ is a constant and can be omitted. 
Working with the modified dissipators $d'_j(\rho)=2 F_j \rho F_j$ shows that
\begin{equation}
    \begin{aligned}
    \left[h_l,F_j\right]=0 &\Rightarrow \left[u_l,d_j\right]=0,\\
    \{h_l,F_j\}=0 &\Rightarrow \{u_l,d_j\}=0.\\
    \end{aligned}
\end{equation}
This means that the (anti)commutation relations of the Hamiltonian terms and jump operators carry on directly to the superoperator level.



Following the above recipe, each of the Lindblad models we study here can be represented with a frustration graph because all its terms (omitting the constant ones) either commute or anti-commute with each other. Fig.~\ref{fig:frustration} depicts frustration graphs of the cluster model with two examples of dissipation channels that we will study: $Y$ and $ZIZ$ jump operators.

There are two useful observations that we can make from the frustration graphs. First, we can spot blockades, which split the graph into independent parts. Second, we can infer an effective model for it.

In Fig.~\ref{fig:frustration}, the Hamiltonian and dissipation terms are represented by red and green vertices respectively. 
Gray vertices correspond to frozen degrees of freedom in a given fragment. Previously, in Section~\ref{sec:worked_examples}, we identified frozen degrees of freedom in the state operator as being the sites of the Pauli string that do not change under the unitary dynamics. Here, in the context of frustration graphs, we identify the same frozen degrees of freedom by the fact that, in the given fragment, the superoperators $u_j$ corresponding to the unitary dynamics of the frozen sites are null, that is they act as 0 in the fragment. 

These vertices can act as \textit{blockades} that separate active regions from each other. For example, if the superorperators describing the dynamics of the model span at most two sites, then a blockade arises if there is a single frozen site, because the dynamical superoperators cannot act on both sides of the blockade at once, making the two subsystems independent. This is illustrated for $ZIZ$ jump operators on \figref{fig:frustration}[b]. However, if the superoperators span three sites, two adjacent frozen sites are necessary in order to disconnect two parts of the system and in this way act as a blockade. 
This is the case with $Y$ jump operators, and \figref{fig:frustration}[a] illustrates that two adjacent frozen sites are needed to fully separate active regions into independent subsystems.
The effective Hamiltonian is then a sum of commuting terms acting on different active regions~\cite{Moudgalya_2022}. 
We refer to this phenomenon as subsystem fragmentation in contrast to the Krylov-subspace fragmentation seen previously. 
The formation of blockades and further splitting of fragments is specific to each model. 
Frustration graphs allow to easily identify subsystem fragments that split the space as illustrated in \figref{fig:steps}[c]. For example, the model in \figref{fig:frustration}[b] is separable into odd and even sublattices, and the odd sublattice is further divided into two pieces by a frozen site, resulting in three independent subsystems, which mathematically translates into the following decomposition: $\mathcal{L} = \mathcal{L}_{\text{odd},1}\otimes\mathds{1}\otimes\mathds{1} + \mathds{1}\otimes\mathcal{L}_{\text{odd},2}\otimes\mathds{1} + \mathds{1}\otimes\mathds{1}\otimes\mathcal{L}_{\text{even}}$.

Working with frustration graphs is also useful to find an effective model for the subsystem fragments -- we can choose any set of Pauli strings that has the same frustration graph. Moreover, following Ref.~\cite{Chapman2020}, it directly informs us, by the presence or absence of forbidden subgraphs, whether the given model is free-fermion solvable.
We build the effective models for our two examples in Section~\ref{sec:Yeffmodel} and~\ref{sec:ZIZeffmodel}. We observe that in the $Y$ case, the graph has a ladderlike structure, which may contain forbidden subgraphs, like the claw highlighted in orange in \figref{fig:frustration}[a]. Such a subgraph prevents the model from being mappable to free fermions and could lead to ergodic fragments.
In the $ZIZ$ case, on the other hand, the frustration graph of each subsystem fragment is a simple chain equivalent to a non-Hermitian Ising model, which is exactly solvable. 

\subsection{Y jump operators}
\label{sec:Yeffmodel}

The jump operators studied here are $F_j=Y_{j}$.
In the basis defined in \eqref{eq:tildebasis1} and \eqref{eq:tildebasis2}, they span three sites in the bulk $F_{j}=-\tx_{j-1}\ty_j\tx_{j+1}$ ($j=2,...,N-1$), and two sites on the boundaries ($F_1=-\ty_1\tx_2$ and $F_N=-\tx_{N-1}\ty_N$). Their spatial extent is what complicates the picture: odd and even sites are not decoupled and at least two frozen sites are needed to separate active regions into disjoint subsystems, as illustrated on \figref{fig:frustration}[a]. 

From the frustration graph in \figref{fig:frustration}[a], one way to write down the effective action of the Lindbladian is as follows.
Each active site can be in two states: $\tx$ or $\ty$, so it behaves like a two-level spin and the size of the effective Hilbert space is equal to $2^{M}$, where $M$ is the number of active sites. 
Therefore, a natural way of writing the effective action of the Lindbladian is to set 
$\tx \equiv \ket{\uparrow}$ and $\ty\equiv \ket{\downarrow}$. In this basis, the action of the $u_l$ and $d_j$ superoperators, shown in Eqs.~\eqref{eq:unitary_action} and \eqref{eq:dissip_action}, becomes:
\begin{equation}
\begin{aligned}
    u_l &\longrightarrow 2 \sigma^y_l,\\
    d_j & \longrightarrow 2 \sigma^z_{j-1}\sigma^z_j\sigma^z_{j+1},
\end{aligned} \label{eq:map1}
\end{equation}
where $\sigma^i$ are the Pauli matrices acting in $\{ \ket{\uparrow} , \ket{\downarrow} \}$. On the boundaries, or when a single site is frozen and does not enter into the effective model, we have instead: $d_j \longrightarrow \pm 2 \sigma^z_j\sigma^z_{j+1}$ or $\pm 2\sigma^z_j$, where the sign $\pm$ depends on the state of the frozen sites.

\begin{figure}[!ht] 
	\centering
	\includegraphics{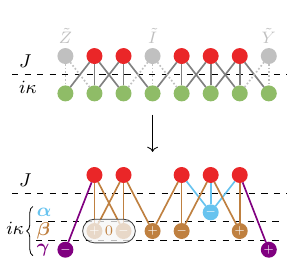}
    \caption{Mapping of the cluster model with $Y$ jump operators ($-\tx\ty\tx$ in the tilde basis) to the effective model in Eq.~\eqref{eq:Y_Leff} using the frustration graph. The state chosen as illustration generates a $2^5$-dimensional fragment (5 active sites). The jump operators give rise to terms spanning three, two or only one active site. These are, respectively, the $\alpha$, $\beta$ and $\gamma$ terms of the effective model. Each term comes with an associated sign derived from the states of the frozen sites. For this reason, in this example, the first two $\beta$ terms interfere destructively and cancel out.}
	\label{fig:steps3}
\end{figure}%

For example, we take a fragment generated by an operator of the form $\tz..\Tilde{I}...\ty$, where dots are active sites. The mapping is explicitly illustrated in \figref{fig:steps3}. Each dissipation term maps to a term spanning one, two or three sites in the effective model, where every term has an associated sign depending on the configuration of frozen sites. It can happen that two such terms map to the same effective action. This is, for example, the case of the second and third dissipation terms (green vertices) depicted in \figref{fig:steps3}, that pick up opposite signs during the mapping and hence cancel out, leading to $\beta_1=0$. The vectors of parameters $\Vec{\alpha}$, $\Vec{\beta}$ and $\Vec{\gamma}$ characterize the three-site, two-site and one-site effective terms, respectively. The effective model in this example with $M=5$ active sites has the following parameters
\begin{equation}
    \begin{aligned}
        \Vec{\alpha} &=(0,0,-1),\\
        \Vec{\beta} &=(0,1,-1,1),\\
        \Vec{\gamma} &=(-1,0,0,0,1).
    \end{aligned}
\end{equation}

In general, therefore, the effective non-Hermitian Lindbladian in this case can contain terms spanning one, two or three sites. 
Ignoring the constant terms, it reduces to
\begin{equation}
\begin{aligned}
    \mathcal{L}_{\text{eff}}&=-2iJ\sum_{l=1}^M \sigma_l^y + 2\kappa \sum_{l=2}^{M-1} \alpha_l \; \sigma^z_{l-1}\sigma^z_{l}\sigma^z_{l+1} \\ &+ 2\kappa \sum_{l=1}^{M-1} \beta_l \; \sigma^z_{l}\sigma^z_{l+1} + 2\kappa \sum_{l=1}^{M} \gamma_l \; \sigma^z_{l},
\end{aligned}\label{eq:Y_Leff}
\end{equation}
where $\alpha_l=0$ or $-1$, $\beta_l=0,\pm 1$ or $\pm 2$, and $\gamma_l=0,\pm 1, \pm 2$ or $\pm 3$ depending on the precise form of the active region and the state of the frozen sites.

In summary, the effective model looses translational invariance because of the fragmentation, it involves interaction terms spanning from one to three sites, and in some cases it cannot be mapped to free fermions. This suggests that it could be chaotic, which is studied further in Sec.~\ref{sec:Y}.

\begin{figure*}[ht!]%
	\centering%
    \includegraphics{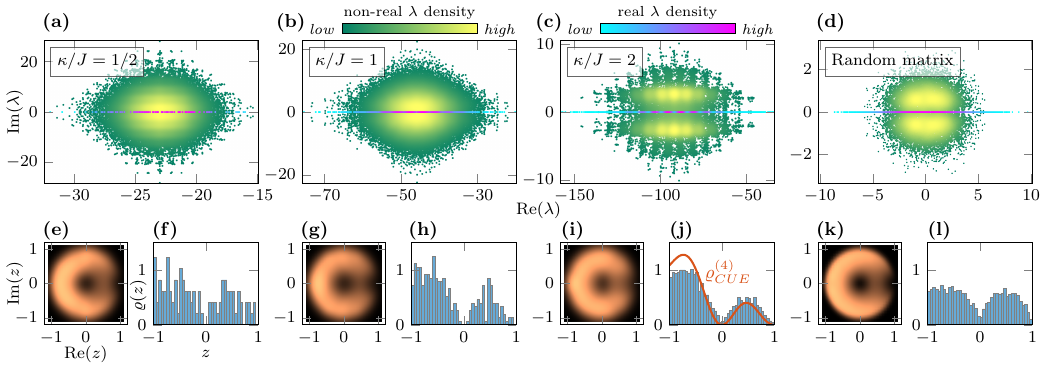}%
	\caption{ (a-c) Spectra of the effective non-Hermitian Hamiltonian model with Y jump operators, of length $15$ and of the form: $\ty\tz......\tz.\Tilde{I}........\tz\ty$, where dots are active sites. (e, g, i) Complex spacing ratios $z$ density for the nonreal part of the spectra in (a-c) respectively. (f, h, j) Histogram of the spacing ratios $z$ for the real part of the spectra in (a-c) respectively. (d) Distribution of the eigenvalues of $2^{13}$ $4\times4$ random matrices drawn from the ensemble defined in \eqref{eq:randensemble}, with $\chi = 2$. (k, l) Ratio statistics for the nonreal and real parts of matrices from the random ensemble ($\chi = 2$), obtained from $70$ matrices of size $2^{15}\times 2^{15}$. Both complex and real spacing ratios were computed by considering $1/3$ of the eigenvalues located in the middle of the spectrum, within an ellipse with axes proportional to the variances of the spectral levels along real and imaginary directions (except for (f), where the whole real part was used due to little number of eigenvalues).}%
	\label{fig:Yjumps}%
\end{figure*}%

\subsection{ZIZ jump operators}
\label{sec:ZIZeffmodel}

In the case of dissipation generated by $F_j=Z_{j-1}Z_{j+1}=\tx_{j-1}\tx_{j+1}$ for $j=2,...,N-1$, every jump operators acts only on odd or even sites. Since the Hamiltonian acts on single sites, we immediately see that odd and even-site chains are independent in each fragment, as shown in \figref{fig:frustration}[b]. Each chain has the same frustration graph as the Ising model. Explicitly, by a change of basis such that the red vertices are the bonds and the green vertices -- the transverse field, the dynamics of the model (restricted to an active subsystem bounded by two frozen sites) are effectively described by a non-Hermitian Hamiltonian of the following form
\begin{equation}
\begin{aligned}
    \mathcal{H}_{\text{eff}}\equiv i\mathcal{L}_{\text{eff}}&=J\sum_{l=1}^{M} \sigma^x_l \sigma^x_{l+1} + i\kappa \sum_{l=2}^{M}\sigma^z_{l} \\&+ i\kappa \left( \zeta_L \sigma^z_1 + \zeta_R\sigma^z_{M+1}\right).
\end{aligned}\label{eq:nh-ising}
\end{equation}
Here, $\zeta_L,R=0$ or $1$, as in some cases, the transverse field terms can be missing from the edges of a given active region of the frustration graph. 

For a region of $M$ active sites, the length of the effective model is $M+1$, but there is a global spin-flip symmetry, $\prod_l \sigma^z_l$, that splits it into two sectors. We only need to consider one of the two sectors depending on the conditions on the boundaries of the active region. More specifically, it depends on the state of the frozen sites at both boundaries of the active region. Appendix~\ref{appendix:formulation} provides a more detailed derivation of the effective model, with a clear explanation of the relationship between the global symmetry and the frozen sites neighbouring the active region. 

To conclude, our model is equivalent to an Ising model with an imaginary field capturing the effect of dissipation on the space of operators.
Various forms of the non-Hermitian Ising model have recently been studied in the context of $PT$-symmetry breaking phase transitions Refs.~\cite{PhysRevA.90.012103,STARKOV2023169268,PhysRevB.104.195137,Yang2022Apr}.

\section{Spectral and dynamical effects}
\label{sec:conseq}

In this section, we study the spectrum and dynamics of Lindbladians with $ZXZ$ Hamilonian with $Y$~\eqref{eq:Y_Leff} and $ZIZ$~\eqref{eq:nh-ising} jump operators and realize universal properties arising due to the underlying physics of non-Hermitian Hamiltonians associated with them.

\subsection{Dissipative quantum chaos}
\label{sec:Y}

For the choice of $Y$ jump operators, the claw subgraphs shown in \figref{fig:frustration}[a] can rule out a free-fermion solution of the effective model. This is determined by the structure of frozen sites in each fragment, even though some special cases might be mappable to free fermions. A generic fragment is not free-fermionic. The particular choice of fragment we study in this section is in fact non-integrable, as it displays chaotic level statistics.
Moreover, model~\eqref{eq:Y_Leff} has  $PT$ symmetry, which arises from the pseudo-Hermiticity with intertwining operator $\eta$, such that $\eta H = H^\dagger \eta$, where $\eta$ is a linear parity, together with the Hermitian conjugation (time reversal). In this case, the parity is the global spin flip: $\eta = \prod_l \sigma_l^z$.
As a result, we obtain fragments with chaotic spectra, but in which the $PT$-symmetry breaking drives a transition from a purely imaginary to a purely real spectrum.

We illustrate the spectral properties by considering a fragment of $15$ active sites, of the form $\ty\tz......\tz.\Tilde{I}........\tz\ty$, where dots represent active sites. \figref{fig:Yjumps}[a-c] show the spectrum at three different parameter regimes. First, we observe that as $\kappa/J$ is increased, the number of purely real eigenvalues increases, in particular past the $\kappa=J$ point. This is a consequence of the $PT$ symmetry, and $\kappa=J$ is the point around which a significant proportion of the eigenvalues on the complex plane are forced onto the real line, where they meet in pairs at exceptional points and are constrained to remain on the real line as $\kappa/J$ is further increased. \figref{fig:frac_ecc}[a] in Appendix~\ref{appendix:choas} shows the fraction of real eigenvalues transitioning from $0$ to $1$.

Second, we compute the distribution of spacing ratios in order to characterize the local properties of the spectrum. The spacing ratio is defined as follows~\cite{PhysRevX.10.021019}
\begin{equation}
    z_k=\frac{\lambda^{NN}_k - \lambda_k}{\lambda^{NNN}_k - \lambda_k},
\end{equation}
where for each eigenvalue $\lambda_k$, $\lambda^{NN}_k$ and $\lambda^{NNN}_k$ are its nearest and next-to-nearest neighbors, respectively, on the complex plane. This quantity is in general complex, except if the considered part of the spectrum is real. In particular, we consider the nonreal (\textit{i.e.}\@ exluding the purely real eigenvalues) and purely real parts of the spectrum separately, as due to the $PT$ symmetry a significant fraction of the spectrum can be constrained to the real line. Within the nonreal part, we compute the spacing ratios for the upper and lower half-planes separately. 
The results are shown on \figref{fig:Yjumps}[e-j] for the spectra in \figref{fig:Yjumps}[a-c]. The heatmaps are the distributions of complex spacing ratios of the nonreal parts of the spectra, while the histograms are the distributions of the spacing ratios on the real line. For each of the spectra presented, both real and nonreal ratio distributions have depleted regions in the middle, which clearly indicates that there is level repulsion.

The properties of local level repulsion in the model are also similar to the random matrices, as can be seen by comparing \figref{fig:Yjumps}[e-j] with \figref{fig:Yjumps}[k-l]. 
We also observe a qualitative resemblance between the spectrum of the model and that of random matrices (compare \figref{fig:Yjumps}[c] and \figref{fig:Yjumps}[d]). 
The significant fraction of levels on the real line creates a \textit{saturn effect} and splits the rest of the spectrum into two lobes on each side. 

The random matrix ensemble referred to here is defined and discussed in Appendix~\ref{appendix:choas} and has been inspired by Refs.~\cite{PhysRevE.94.012147,PhysRevE.96.012154}. It is an ensemble of real, symmetric matrices which are also made $PT$-symmetric/pseudo-Hermitian with the same intertwining operator $\eta$ as our physical model, and in which a parameter $\chi$ controls the relative strength of Hermitian and anti-Hermitian components, similarly to $\kappa/J$. 

In conclusion, we have learned that the effective model displays level repulsion in its spectrum, which is confirmed by its similarity to random matrices. However, we claim that there exists a transition related to the $PT$ symmetry of the model, especially visible in the fraction of real eigenvalues (\figref{fig:frac_ecc}[a]). The random ensemble that we have introduced in Appendix~\ref{appendix:choas} has a $PT$ symmetry as well, but it does not capture the transition in the thermodynamic limit (see \figref{fig:frac_ecc}[b]). Therefore, the universality classes of the model and of the random matrices are likely to be different. Although the full behavior is not described by this ensemble, it is the most accurate among other ways to construct pseudo-Hermitian matrices. These can be found for example in Refs.~\cite{PhysRevE.67.045106,Birchall_2012,Bohigas2013,Graefe_2015,PhysRevResearch.6.023303,Pato_book}. Refs.~\cite{Feinberg_2021,Feinberg_2022} also introduce a parameter controlling the fraction of real eigenvalues, but the associated spectra have a completely different appearance.


\subsection{Integrability and criticality}
\label{sec:ZIZ}

\begin{figure}[!ht] 
	\centering
    \includegraphics{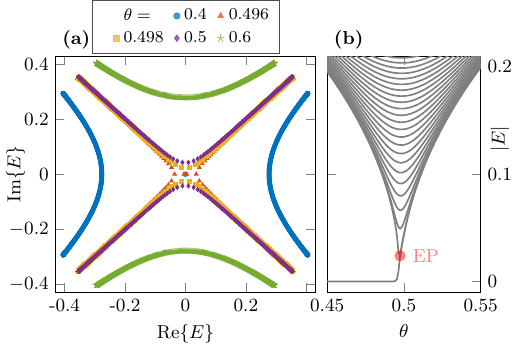}
	\caption{Properties of the non-Hermitian Hamiltonian~\eqref{eq:nh-ising}. (a) Complex spectrum (of the corresponding Majorana fermions) for different parameters set by $\theta$, such that $J=\cos(\theta \pi/2)$ and $\kappa=\sin(\theta \pi/2)$. (b) Absolute value of the energy, displaying the gap closing and the zero mode vanishing as $\theta$ increases, with an exceptional point (EP) highlighted in red. The size of the system is $M+1=400$ in (a) and (b).}
	\label{fig:ZIZ}
\end{figure}%

The non-Hermitian Ising Hamiltonian with boundary fields~\eqref{eq:nh-ising} is solved using the standard Jordan-Wigner transformation to free fermions, further decomposed into Majorana fermions. The complex spectrum of the corresponding free Majorana fermions is shown in \figref{fig:ZIZ}[a]. When $\kappa<J$, the spectrum contains a zero mode localized on the edges of the system. \figref{fig:ZIZ}[b] shows that there is a transition at around $\kappa=J$, beyond which the zero mode disappears. All the necessary details for an exact solution and understanding of the model can be found in Appendix~\ref{appendix:solution}.

Highlighted in red in \figref{fig:ZIZ}[b] is an exceptional point (EP), which in the thermodynamic limit coincides with the phase transition of the Ising model. As such, the exact solution found here allows us to relate the exceptional points in the spectrum of the original Lindbladian to critical points of the effective model and to associated topological properties. The physical meaning of EPs is reviewed in Ref.~\cite{UedaNH}. Moreover, they can be probed by operator dynamics. Non-Hermitian phase transitions, non-equilibrium dynamical phases and EPs can be detected and studied with the Loschmidt echo in the quantum dynamics after a quench~\cite{agarwal2022,PhysRevResearch.3.023022,PhysRevB.109.094311,Tang_2022,PhysRevB.100.184313}.

We demonstrate the signature of EPs in the Lindblad dynamics. 
by computing a specific two-time correlation function of an operator $O$.
An initial state is prepared by measuring a Hermitian traceless operator $O$ which projects the system into $\rho_0 = \mathds{1}\pm O$. Following the Lindbladian evolution for time $t$, operator $O$ is measured again:
\begin{align}
    \mathcal{E}
    &= \Tr(Oe^{\mathcal{L}t}(\mathds{1}\pm O)) = \Tr(Oe^{\mathcal{L}t}\mathds{1}) + \Tr(Oe^{\mathcal{L}t}O) \nonumber \\&= \Tr(Oe^{\mathcal{L}t}O).
\end{align}
In the effective picture, $O(t)$ corresponds to a state $\ket{O(t)}$ and the above quantity is the Loschmidt echo
\begin{equation}
    \mathcal{E} = \braket{O(0)}{O(t)} 
\end{equation}
In practical calculations, the state evolved by non-unitary dynamics generated by the Hamiltonian~\eqref{eq:nh-ising} is renormalized at each instant of time, and the quantity of interest is
\begin{equation}
    \label{eq:echo}
    \tilde{\mathcal{E}} = \frac{\braket{O(0)}{O(t)}}{\sqrt{\braket{O(t)}{O(t)}}}. 
\end{equation}

\begin{figure}
    \centering
    \includegraphics{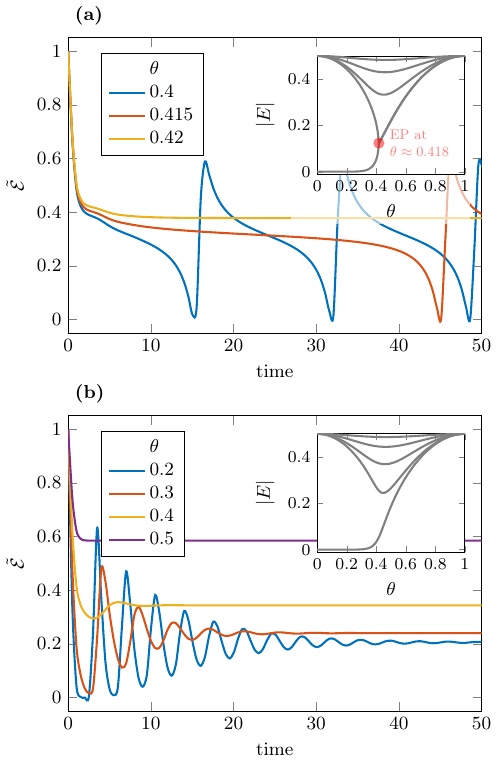}
    \caption{Loschmidt echo $\tilde{\mathcal{E}}$ of the all-up state $\ket{O(0)} = \ket{\uparrow\uparrow\ldots\uparrow}_{\sigma^z}$. Fragments with effective models of (a) $M+1=8$ sites, and (b) $M+1=9$ sites. The EPs are present only in fragments with an odd number of active sites $M$, which has a dramatic effect on quench dynamics. The insets show the absolute value of the energy and the gap closing as $\theta$ (see \figref{fig:ZIZ} for the definition) is varied across the transition.}
    \label{fig:echo}
\end{figure}

As the initial state, we choose the all-up state along the $z$-direction:
\begin{equation}
     \ket{O(0)} = \ket{\uparrow\uparrow\ldots\uparrow}_{\sigma^z}. \label{eq:allup}
\end{equation}
The results of the quench dynamics are shown in \figref{fig:echo}. First, the insets show that models with an odd number $M$ of active sites have EPs in their spectrum (\figref{fig:echo}[a]), while even-$M$ models do not (\figref{fig:echo}[b]). Second, we observe that this has significant consequences for the Loschimdt echo after a quench. In the presence of an EP, the echo continues to oscillate in time for all values of $\theta$ below the transition. As $\theta$ above the EP, the echo monotonically decays to a finite value. The transition between the two regimes is abrupt, and appears as a first-order transition. On the other hand, in the absence of EPs, the echo continuously transitions from an underdamped to an overdamped regime. 

The initial state~\eqref{eq:allup} in the effective model corresponds to a simple operator in the original Lindblad dynamics. For example, if the model describes a subsystem fragment of 3 active sites on the odd sublattice, the operator $O$ can be written as
\begin{equation}
    O = {\color{gray!50}\ldots} \;{\color{gray}\ti} \; {\color{gray!50}\ti} \; {\color{myred}\tx}\; {\color{gray!50}\ti}\; {\color{myred}\tx}\; {\color{gray!50}\ti} \; {\color{myred}\tx}\; {\color{gray!50}\ti} \; {\color{gray}\ti} \; {\color{gray!50}\ldots}, \label{eq:originalO}
\end{equation}
where the light gray even sites do not matter and have been set to identity, and the extremities are sites frozen in the identity state, delimiting the active region from the rest (Appendix~\ref{app:corresp} provides further clarification of this correspondence). In this manner, we see that non-trivial non-Hermitian features can be accessed through the Lindblad dynamics of relatively simple operators.

Our analytical framework has unravelled the fragmented operator space of the Lindblad superoperator and the hidden integrability of the fragments is distinct from the properties of the Hamiltonian and the jump operators. Furthermore, this tool allowed to relate the properties of the open system to an effective non-Hermitian model and in particular established a link between the EPs and non-Hermitian criticality. Alternatively, these results can be viewed as a way of engineering and probing the properties of a non-Hermitian Hamiltonian through Lindbladian dynamics of open systems, namely by studying EPs through the Loschmidt echo and quench dynamics.


\section{Discussion and outlook}
\label{sec:conclusion}
In this work, we have formulated the theory of operator-space fragmentation for Lindbladian dynamics generated by Pauli-string operators. By generalizing the definitions of bond and commutant algebras, describing Hilbert-space fragmentation, to the superoperator level, we have provided a unified framework for fragmentation of Lindbladions with both weak and strong symmetries. This framework reveals how the interplay between unitary dynamics and dissipation can lead to the emergence of exponentially many Krylov subspaces in the operator space, each hosting distinct dynamical regimes--from integrability to chaos.

For concreteness, we have studied a family of models consisting of frustration-free Hamiltonians and Pauli-string jump operators. These models exhibit a hierarchy of fragmentation: first into \textit{subspace fragments} determined by conserved patterns of frozen sites, and further into \textit{subsystem fragments} dynamically decoupled by blockades. This mirrors Hilbert-space fragmentation in closed systems but extends its scope to the dissipative dynamics of mixed states.

To analyze the dynamics within individual fragments, we introduced the use of frustration graphs for open quantum systems. These graphs summarize the algebraic properties of the Lindbladian restricted to a single fragment and allow us to map the dynamics to effective non-Hermitian Hamiltonians. For example, in the case of $ZIZ$-type dissipation, the frustration graph reveals that each fragment maps to a non-Hermitian transverse-field Ising model, which is exactly solvable via Jordan-Wigner transformation. The spectrum of this effective model exhibits exceptional points (EPs) around $\kappa = J$, corresponding to a dissipative phase transition. These EPs, which coincide with the gap closing in the Ising-like fragments, provide a direct link between the Lindbladian spectrum and non-Hermitian critical phenomena. Moreover, this can be observed physically due to the connection between two-time correlations and the Loschmidt echo that we have highlighted.

In contrast, for $Y$-type dissipation, the frustration graph precludes a free-fermion mapping, leading to chaotic fragments. The spectrum of these fragments exhibits a mix of chaotic behavior and structures emerging due to $PT$ symmetry. Specifically, as the dissipation strength $\kappa$ increases relative to the unitary coupling $J$, the spectrum undergoes a transition from a regime dominated by purely imaginary eigenvalues to one dominated by purely real eigenvalues. This transition is diagonozed by level repulsion and random-matrix statistics, characteristic of quantum chaos, in parts of the spectrum  while the $PT$ symmetry imposes constraints that prevent the spectrum from being fully ergodic. This interplay between chaos and symmetry highlights the rich dynamical behavior that can emerge in open quantum systems, and which can be accessed and studied using the set of tools introduced in this article.

In summary, our work unifies the physics of open quantum systems and non-Hermitian Hamiltonians by highlighting the role played by operator space fragmentation which shows a new form of hierarchical structure involving both operator and real space characteristics. The universal dynamical properties such as quantum chaos and integrability for dissipative models are realized which undergo novel forms of dynamical phase transition. By uncovering hidden structures in the operator space, our work provides a framework for discovering new dynamical phenomena in open quantum systems.

The class of models we have investigated serves as a versatile playground for constructing effective non-Hermitian Hamiltonians in open quantum systems. Engineered dissipation offers an opportunity to realize a wide range of dynamical behaviors in the fragments and novel forms of non-equilibrium quantum order.
Beyond its physical implications, our framework provides a practical tool for accessing free-fermionic solvability of an open quantum system. This method, rooted in the algebraic structure of the Lindbladian and the connectivity of its frustration graphs, is easily applicable to all frustration-free Hamiltonians with Pauli-string jump operators. It offers a systematic way to identify free-fermion solvable models and predict their spectral properties. Generalization of this framework to higher dimensions and non-local graphs would open opportunities to study a wider class of models.

Finally, the fragmentation of the operator space has significant consequences for the dynamics of quantum information. By decoupling the evolution of operators into distinct subspaces, fragmentation hinders operator spreading and restricts the flow of information within the system. This decoupling effect arises from the conserved patterns of frozen sites and the dynamical blockades, which isolate subsystems and prevent the system from exploring the full Hilbert space. The restricted flow of information can provide novel mechanisms for protecting information by limiting the noise in the logical space and thereby efficiently implement quantum error correction with low overheads.

\begin{acknowledgements}
A.P.\ and D.C.R. are funded by the European Research Council (ERC) under the EU’s Horizon 2020 research and innovation program (Grant Agreement No.~853368).
C.J.T. is supported by an EPSRC fellowship (Grant Ref. EP/W005743/1).
D.P.\ is funded by the UCL Graduate Research Scholarship. 
The authors acknowledge the use of the UCL Myriad High Performance Computing Facility (Myriad@UCL) and of the High Performance Computing cluster in the London Centre for Nanotechnology, and associated support services, in the completion of this work.
\end{acknowledgements}


\appendix

\section{$PT$-symmetric random matrix universality}
\label{appendix:choas}

In this section, we highlight and provide further evidence of random matrix properties of model~\eqref{eq:Y_Leff} with $Y$-jumps as discussed in Sec.~\ref{sec:Y}. We introduce a random matrix ensemble with the appropriate symmetries to analyze the classify the RMT universality of the model and its deviations away from it. 
We are focussed on two essential spectral properties~\cite{PhysRevE.94.012147,PhysRevE.96.012154}: the fraction $f_r$ of purely real eigenvalues, and the eccentricity $\varepsilon$ of the spectrum.
The eccentricity is defined by assigning an ellipse to the spectrum, excluding the purely real eigenvalues. The minor and major axes, respectively called $a$ and $b$, are determined from the standard deviations of the distribution of these eigenvalues, along the real and imaginary axes. The eccentricity can therefore be defined as $$\varepsilon=\sqrt{1-\left(b/a\right)^2}.$$

\begin{figure}
    \centering
    \includegraphics{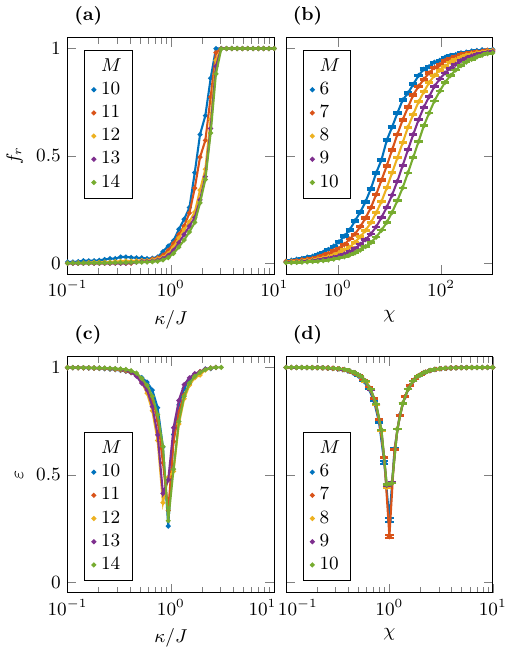}
    \caption{(a) Fraction of real eigenvalues $f_r$ as a function of $\kappa/J$ for a specific choice of the effective model~\eqref{eq:Y_Leff} with $M$ active sites. (b) Average $f_r$ in random matrices of size $2^M$ by $2^M$ drawn from the ensemble defined in Eq.~\eqref{eq:randensemble}, as a function of parameter $\chi$. (c, d) Eccentricity $\varepsilon$ of the spectrum (excluding purely real eigenvalues) for the same model and random matrices as in (a, b), respectively. The precise form of the models with different $M$ are given by: $\ty\tz....\tz.\Tilde{I}.....\tz\ty$, $\ty\tz....\tz.\Tilde{I}......\tz\ty$, $\ty\tz.....\tz.\Tilde{I}......\tz\ty$, $\ty\tz.....\tz.\Tilde{I}.......\tz\ty$, $\ty\tz.....\tz.\Tilde{I}........\tz\ty$.}
    \label{fig:frac_ecc}
\end{figure}

In \figref{fig:frac_ecc}[a,c], we show how these two quantities, $f_r$ and $\varepsilon$ change as a function of $\kappa/J$. 
We observe that as $\kappa/J$ is increased, there is a smooth transition from a purely imaginary to a purely real spectrum. For $\kappa/J \ll 1$, the scaling of $f_r$ rapidly tends to zero, while for $\kappa/J \gg 1$ the real fraction is one. For intermediate values of $\kappa/J$, the real fraction smoothly crosses over between the two limits for finite size systems. The scaling seems to suggest that the crossover regime may survive in the thermodynamic limit, instead of a sharp jump from zero to one.

We compare these observations to the behavior of random matrices. The random matrix ensemble we consider is defined to capture the symmetries of the effective model. In analogy with the non-Hermitian Hamiltonian, the random matrix model is required to be real, pseudo-Hermitian and symmetric. We would also like to define an ensemble where a parameter, $\chi$, can drive a transition between a purely real and a purely imaginary spectrum, as it is observed in our model. A matrix $A$ belonging to such a random ensemble is obtained as follows.
We draw a random real matrix $X \in \mathds{R}^{n\times n}$ with entries chosen independently following the normal distribution. This matrix is then symmetrized according to: $H=\left(X+X^T\right)/2$. Finally, we transform it to being pseudo-Hermitian, \textit{i.e.}\@ satisfying $\eta H = H^\dagger \eta$, where $\eta$ is a linear parity, in our case in the main text, $\eta = \prod_l \sigma_l^z$.
For implementing this, we use projectors $P$ and $Q$ onto positive and negative eigenvalue sectors of $\eta$, such that $\eta=P-Q$. The random matrix $A$ possessing the desired properties is then obtained as~\cite{PhysRevE.94.012147,PhysRevE.96.012154}
\begin{equation}
    A = \chi \left(PHP + QHQ \right) + \left( PHQ - QHP \right),
    \label{eq:randensemble}
\end{equation}
where the first two terms are the Hermitian blocks and the last two terms are the anti-Hermitian interactions between them. The parameter $\chi$ regulates the strength of the Hermitian blocks relative to the anti-Hermitian off-diagonal blocks. It is akin to $\kappa/J$ of our model~\eqref{eq:Y_Leff}, which sets the strength of Hermitian versus anti-Hermitian terms.

\figref{fig:frac_ecc} compares the model and the random matrix ensemble, in terms of the fraction of real eigenvalues and the eccentricity of the spectrum. 
The effect of $\chi$ in \figref{fig:frac_ecc}[b, d] is qualitatively similar to the effect of $\kappa/J$ in \figref{fig:frac_ecc}[a, c] in forcing a transition between a purely imaginary to a purely real spectrum. 
However, there are also significant differences. 
First, in \figref{fig:frac_ecc}[a] the transition to a purely real spectrum is sharp and happens at a finite $\kappa/J$, while in \figref{fig:frac_ecc}[b], such a sharp transition cannot be seen, there are always some complex eigenvalues for any finite $\chi$. Also, while the position of the crossover is not significantly affected by increasing $M$ in the model, it does consistently drift towards higher $\chi$ in the random ensemble. Naturally, it is expected that the spectrum of the random ensemble follows the elliptic law~\cite{Girko1985,elliptic2014} in the thermodynamic limit, i.e. it tends to a uniform distribution on an ellipse in the complex plane.
Second, the dip of the eccentricity towards 0 in the model (\figref{fig:frac_ecc}[c]) tends to happen for $\kappa/J<1$, while it happens exactly at $\chi=1$ in the random ensemble (\figref{fig:frac_ecc}[d]). This is due to the fact that in the model~\eqref{eq:Y_Leff}, if some sites in the bulk of the fragment are frozen, there are fewer anti-Hermitian terms than Hermitian ones. 



\section{From frustration graph to the non-Hermitian TFIM}
\label{appendix:formulation}

In Sec.~\ref{sec:subfrag}, we showed the mapping of the Lindblad master equation~\eqref{eq:lindblad2} of the cluster model with $ZIZ$ jump operators to a non-Hermitian transverse field Ising model (TFIM) of the form~\eqref{eq:nh-ising}. This requires a few intermediate steps that we present in  detail here. 

\subsection{A natural basis: intermediate model}
\label{appendix:naturalbasis}

Each subsystem fragment of the operator space is characterized by a pattern of frozen and active sites, as shown in Sec.~\ref{sec:cluster}. Each active site can be in two states: $\tx$ or $\ty$, so it behaves like a two-level spin and the size of the effective Hilbert space is equal to $2^{M}$, where $M$ is the number of active sites. 

Therefore, a natural way of writing the effective action of the Lindbladian is to set 
$\tx \equiv \ket{\uparrow}$ and $\ty\equiv \ket{\downarrow}$. In this basis, the action of the $u_l$ and $d_j$ superoperators, given in Eqs.~\eqref{eq:unitary_action} and \eqref{eq:dissip_action}, transforms as follows:
\begin{equation}
\begin{aligned}
    u_l &\longrightarrow 2 \tau^y_l,\\
    d_j & \longrightarrow 2 \tau^z_{j-1} \tau^z_{j+1},   
\end{aligned} \label{eq:map1}
\end{equation}
where $\tau^i$ are the Pauli matrices acting in $\{ \ket{\uparrow} , \ket{\downarrow} \}$. 

\begin{figure}[!ht] 
	\centering
	\includegraphics{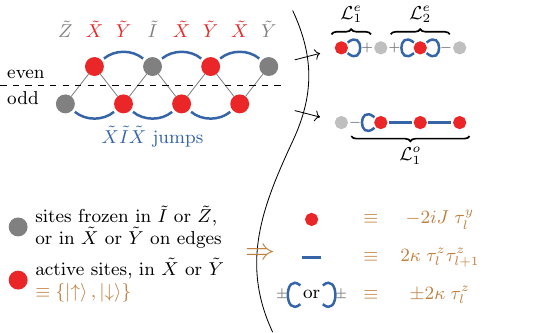}
	\caption{Illustration of subsystem fragmentation and mapping to the effective model in Eq.~\eqref{eq:nh-ising_interm}. Cluster model with $ZIZ$ jumps ($\tx\Tilde{I}\tx$ in the tilde basis), with a state generating a $2^5$-dimensional fragment (5 active sites).}
	\label{fig:steps2}
\end{figure}%
 
To understand the transformation concretely, consider
for example the fragment corresponding to a lattice of $8$ spins with sites $1$, $4$ and $8$ frozen, as depicted in Fig.~\ref{fig:steps2}.
The odd and even sublattices decouple. Moreover, $\mathcal{L}$ in this fragment acts independently on three disjoint supports: $\mathcal{L}=\mathcal{L}^o_1\otimes \mathds{1}\otimes \mathds{1} + \mathds{1}\otimes \mathcal{L}^e_1\otimes \mathds{1} + \mathds{1}\otimes \mathds{1}\otimes \mathcal{L}^e_2$. In the bulk of those active regions the model consists of interactions between neighbouring sites and a transverse field, as given by the mapping~\eqref{eq:map1}.

On the boundaries of active regions, however, there is a subtlety: in the odd chain of the present example, $\mathcal{L}_1^o$, the left edge of the active region has a frozen neighbour, so the dissipator acting on both of them will only effectively act as $\tau_1^z$ on the edge of the active chain, but it will also pick up a sign depending on the state of the frozen neighbour. 
As such, the effective action on the edges can, in general, be written as $\pm2\kappa\tau_1^z$.  
In Fig.~\ref{fig:steps2}, this fact is represented as a blue loop on the edges, accompanied by the corresponding $\pm$ sign.
Note that the other boundary does not have a frozen neighbour, so such a contribution does not appear there.
Ignoring the constant terms, the effective model in this subsystem is
\begin{equation}
    \mathcal{L}_1^o=-2iJ\sum_{l=1}^3 \tau^y_l +2\kappa \sum_{l=1}^2 \tau_l^z\tau_{l+1}^z - 2\kappa \tau_1^z.
\end{equation}
For the even lattice, the reasoning is similar but a frozen site splits it into two single-site subfragments.

In this manner, we can write the effective model in a generic subsystem fragment of $M$ consecutive active sites on the even or odd sublattice as
\begin{equation}
\begin{aligned}
    \mathcal{L}_{\text{eff}}&=-2iJ\sum_{l=1}^{M} \tau^y_l + 2\kappa \sum_{l=1}^{M-1} \tau^z_l \tau^z_{l+1} \\  & +  2\kappa\; \left(\epsilon_L\; \tau^z_1 + \epsilon_R\; \tau^z_M \right) \\ & -2\kappa \left(M+|\epsilon_L|+|\epsilon_R|-1\right)I\\
    &\equiv -i\mathcal{H}_{\text{eff}},
\end{aligned} \label{eq:nh-ising_interm}
\end{equation}
where $\epsilon_{L}=1$ if the leftmost site of the subsystem fragment is adjacent, on its left, to a frozen site in the state $\ti$ (or $\tx$ if the frozen site is a physical boundary), $\epsilon_{L}=-1$ when that frozen site is $\tz$ (or $\ty$ if it is a physical boundary), and similarly for $\epsilon_{R}$. It is also possible for $\epsilon_L$ to be equal to $0$ on the effective model for an even-site subchain, and similarly for $\epsilon_R$ for a model on the odd subchain, as it was the case in the above illustration. Only one of them can be zero at once and this happens when the first or last site of the effective model corresponds, respectively, to the second or second-to-last site of the full original physical chain.


The model is thus equivalent to the Ising model with an imaginary transverse field and longitudinal field on the boundaries. This formulation of the model is, however, not the prettiest and easiest to solve due to the longitudinal fields, which we deal with in the following two subsections. 

\subsection{Extended model to resolve the longitudinal fields}

The model~\eqref{eq:nh-ising_interm} can be solved exactly by generalizing the method of Ref.~\cite{Campostrini_2015} to the non-Hermitian case. The method consists in extending the chain by adding one site on each side, that is at positions $0$ and $M+1$, and defining the extended model (ignoring the constant terms of~\eqref{eq:nh-ising_interm})
\begin{equation}
\begin{aligned}
    \mathcal{H}_{\text{ext}}&=2J\sum_{l=1}^{M} \tau^y_l + 2i\kappa \sum_{l=1}^{M-1} \tau^z_l \tau^z_{l+1} \\&+ 2i\kappa \left( \zeta_L \tau^z_0 \tau^z_1 + \zeta_R \tau^z_M \tau^z_{M+1}\right).
\end{aligned}\label{eq:nh-ising_ext}
\end{equation}
Here $\zeta_{L/R}=0,1$ indicate whether there was a longitudinal field or not at the left or right edge. The crucial observation is that the newly added sites are not subjected to the transverse field, so $\tau^z_0$ and $\tau^z_{M+1}$ are symmetries. The corresponding symmetry sectors are labeled by $(s_0,s_{M+1})$, with $s_0=\pm1,s_{M+1}=\pm1$. In each sector, we recover the model with a given configuration of the boundaries. For example, the models with aligned and anti-aligned edge fields are respectively found in the $(1,1)$ and $(1,-1)$ sectors.

\subsection{Duality transformation to get the final form of the model}

Finally, the Hamiltonian form~\eqref{eq:nh-ising} is obtained by a Krammers-Wannier duality transformation~\cite{PhysRev.60.252} of $\mathcal{H}_{\text{ext}}$, by defining a dual lattice as follows
\begin{equation}
\begin{aligned}
     \sigma^z_l&=\tau^z_{l-1}\tau^z_l,\\
     \sigma_l^x&=\prod_{j\geq l}^{M+1}\tau^y_j,
\end{aligned} \label{eq:map2}
\end{equation}
where $l=0...M+1$. For $l=0$ we need to define $\sigma^z_0=\tau^z_0$ instead. The Hamiltonian then becomes
\begin{equation}
\begin{aligned}
    \mathcal{H}_{\text{eff}}&=2J\sum_{l=1}^{M} \sigma^x_l \sigma^x_{l+1} + 2i\kappa \sum_{l=2}^{M}\sigma^z_{l} \\&+ 2i\kappa \left( \zeta_L \sigma^z_1 + \zeta_R\sigma^z_{M+1}\right).
\end{aligned}
\end{equation}
Ignoring the global factor of $2$ we obtain~\eqref{eq:nh-ising}. We observe that it has possibly missing boundary field terms, $\zeta_{L/R}=0,1$, depending on the boundary conditions of the physical system.

The two symmetries of the extended model~\eqref{eq:nh-ising_ext} determined which symmetry sector of the model we have to choose depending on the sign introduced by the frozen sites neighbouring the described active region. These symmetries here correspond to $\tau^z_0 = \sigma^z_0$ and $\tau^z_{M+1}=\prod_{j=0}^{M+1}\sigma^z_{j}$. 

The $0$-th site does not appear in the model, so we can ignore it. Therefore, for a subsystem fragment consisting of $M$ active sites, we can work with an effective theory on a spin chain of length $M+1$. The remaining global spin-flip symmetry $\prod_{j=1}^{M+1}\sigma^z_{j}$ corresponds to $\tau^z_0 \tau^z_{M+1}$, which indicates that we should look at its $+1$ sector if the signs due to the left and right neighbouring frozen sites are the same, or at the $-1$ sector if they are opposite.

\subsection{Correspondence between the final effective model and the physical system} \label{app:corresp}

To sum up, the series of mappings, Eqs.~\eqref{eq:map1} and~\eqref{eq:map2}, that takes us from the physical model to the non-Hermitian TFIM is
\begin{equation}
\begin{array}{lcl}
     u_l \longrightarrow & 2\tau^y_l & \longrightarrow 2\sigma^x_{l-1}\sigma^x_{l+1},\\
     d_j \longrightarrow & 2\tau^z_{j-1}\tau^z_{j+1} & \longrightarrow 2\sigma^z_{j}.
\end{array} \label{eq:map3}
\end{equation}
The intermediate step of the mapping can be omitted, but it is useful to connect the properties of the final effective model to the physical system. In particular, the final Hilbert space dimension is bigger than the original one, and thanks to the intermediate model, we clearly see which symmetry sector it should be restricted to in order to match given physical boundary conditions.

From these mappings, we can also deduce how states transform. Note that states in the effective model correspond to operators in the Lindbladian picture. 
In particular, in Sec.~\ref{sec:ZIZ}, we propose a quench experiment that could be used to probe effective non-Hermitian physics, and specifically a transition associated to the closing of a spectral gap and EPs. Therefore, it is necessary to know what the initial state undergoing the quench, as well as the measured observables, look like in the original system. 

The initial state we choose is~\eqref{eq:allup}. It is a tensor product of $M+1$ single-spin up states -- the $0$-th site is not included. 

In order to work out how it transforms back to the original basis, we write down a complete set of operators that stabilize the state, transform them, then look for the state which they stabilize in the target basis. 

State~\eqref{eq:allup} is stabilized by $\sigma^z_{l}\; (l=1...M)$, and by the global symmetry $\prod_{j=1}^{M+1}\sigma^z_{j}$, whose only task is to fix the boundary conditions of the model (here we chose the state to be in the $+1$ sector), as explained earlier. 
We only need to consider the $M$ $\sigma^z_{l}$ stabilizers. 
It is straightforward to map them back to their superoperator form by using~\eqref{eq:map3}.

Note that in~\eqref{eq:map3}, the indexing is with respect to the full spin chain, including both odd and even sites, while the state we consider here is only defined on the odd or even sublattice.

Finally, the operator to be quenched is readily obtained as the operator stabilized by the corresponding $d_j$ superoperators. For an example of three active sites ($M=3$), it looks like~\eqref{eq:originalO}. Note that 
\begin{equation}
    O = {\color{gray!50}\ldots} \;{\color{gray}\tz} \; {\color{gray!50}\ti} \; {\color{myred}\ty}\; {\color{gray!50}\ti}\; {\color{myred}\ty}\; {\color{gray!50}\ti} \; {\color{myred}\ty}\; {\color{gray!50}\ti} \; {\color{gray}\tz} \; {\color{gray!50}\ldots}
\end{equation}
is also a valid choice and gives the same dynamics. The difference in these two possible states stems from the frozen sites at the boundaries. This differentiating information in the effective model would be given by the sign of the $\sigma^z_{0}$ stabilizer, but the $0$-th site is not included as it is a symmetry and its two sectors are degenerate.


\section{Detailed solution of the non-Hermitian effective model}
\label{appendix:solution}

\subsection{Bulk spectrum of the non-Hermitian TFIM}

The bulk spectrum of the model \eqref{eq:nh-ising} is studied with periodic boundary conditions (PBC). Although the fragments require open boundary conditions (OBC), the periodic system is easily solvable and helps to solve the open one. It is equivalent to a non-Hermitian Kitaev model in the fermionic formulation. Refs.~\cite{Rahul2022Apr,PhysRevA.94.022119,PhysRevB.97.115436,PhysRevB.104.205131,Shi2023Apr} study phase transitions and topology in that model with possibly different sources of non-Hermiticity, such as imbalanced hopping and supraconducting terms.

Applying the Jordan-Wigner transformation $a_l=\frac{1}{2}(\sigma^x_l+i\sigma^y_l)\prod_{j=1}^{l-1}\sigma^z_j$, the model~\eqref{eq:nh-ising} with PBC (and $\zeta_{L/R}=1$) maps onto
\begin{equation}
    \mathcal{H}= J \sum_{l=1}^{M-1} (a_l^\dagger-a_l)(a_{l+1}^\dagger+a_{l+1}) + i\kappa\sum_{l=1}^M (1-2a_l^\dagger a_l).
    \label{eq:nh-kitaev}
\end{equation}

Then rewriting in terms of Majorana fermions $b_l=a_{l}^\dagger + a_{l}$ and $c_l=i(a_{l}^\dagger - a_{l})$, we get a simpler form
\begin{equation}
    \mathcal{H}=- iJ \sum_{l=1}^{M-1} c_l b_{l+1} + \kappa\sum_{l=1}^M b_lc_l.
\end{equation}

The bulk spectrum can be obtained considering this model with periodic boundary conditions (PBC), analogously to Kitaev's original work~\cite{Kitaev2001Oct}. However, periodic boundary conditions in spins correspond to either periodic or antiperiodic boundary conditions in fermions, depending on the parity of fermions.  With PBC in the spin model, the additional term of the form $\sigma^x_M \sigma^x_1$ transforms to $P_{N_{tot}}(a_M^\dagger - a_M)(a_1^\dagger + a_1)$, where $P_{N_{tot}}=(-1)^{\sum_l a^\dagger_la_l}=e^{i\pi\sum_l a^\dagger_la_l}=\prod_{l}(1-2a^\dagger_la_l)$ is the fermion-number-parity, $N_{tot}$ being the total number of fermions. 
Note that it is conserved by \eqref{eq:nh-kitaev}, such that the dynamics will not mix the two subspaces with different parity.
It corresponds to the spin-flip symmetry $\prod_{l}\sigma^z_l$ of the original model. Therefore, to make the mapping correct, we need to impose appropriate boundary conditions: the $-1$ spin-flip sector maps to the fermionic Hamiltonian with PBC ($a_{M+1}=a_1$) and odd $N_{tot}$, while the $+1$ spin-flip sector maps to the fermionic Hamiltonian with antiperiodic boundary conditions (APBC; $a_{M+1}=-a_1$) and even $N_{tot}$.

\begin{figure}[h] 
	\centering
	\includegraphics{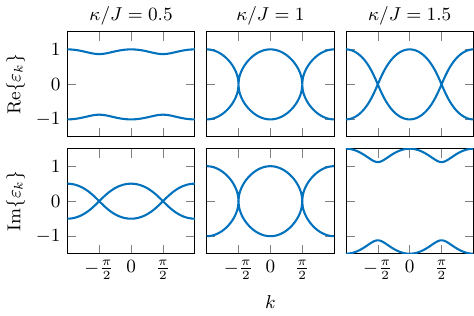}
	\caption{Real and imaginary parts of the complex energy bands of the non-Hermitian TFIM \eqref{eq:nh-kitaev}, for different values of $\kappa/J$, with $J=1$.}
	\label{fig:PBCspec}
\end{figure}

The Fourier transform is defined as $a_l = \frac{1}{\sqrt{M}} \sum_k e^{ikl}a_k$, where the sum is over the first Brillouin zone (BZ). The boundary conditions determine which values the discrete momentum $k$ can take:
\begin{equation}
	\begin{array}{ll}
	   k&=\frac{2\pi m}{M},\; m=0...M-1 \mbox{ (odd }N_{tot}\mbox{, PBC)} \\
        & \\
		k&=\frac{2\pi (m+1/2)}{M},\; m=0...M-1 \mbox{ (even }N_{tot}\mbox{, APBC)}
	\end{array}
\end{equation}
Eq.~\eqref{eq:nh-kitaev} in reciprocal space is then
\begin{equation}
    \mathcal{H}=i\kappa M + 2 \sum_k \psi_k^\dagger H_k \psi_k,
\end{equation}
where we have grouped the positive and negative momentum degrees of freedom in the spinor $\psi_k=\begin{pmatrix}
    a_k \\ a^\dagger_{-k}
\end{pmatrix}$ - the $k$-sum thus running over half of the BZ - to write the Hamiltonian in the Bogoliubov-de Gennes form with
\begin{equation}
    H_k=\begin{pmatrix}
    (J \cos(k)+i\kappa) & iJ \sin(k)\\ -iJ \sin(k) & -(J \cos(k)+i\kappa)
\end{pmatrix}.
\end{equation}
Equivalently, $H_k=y_k \tau^y + z_k \tau^z$ where $\tau^i$ Pauli matrices and $y_k=-J \sin(k)$, $z_k=(J \cos(k)+i\kappa)$. 
The energy levels are $\pm\varepsilon_k$ with
\begin{equation}
    \varepsilon_k= \sqrt{(J \cos(k)+i\kappa)^2 + (J \sin(k))^2}, \label{eq:PBCenergy}
\end{equation}
and are shown on Fig.~\ref{fig:PBCspec} for different values of the parameters.

The corresponding matrix of eigenvectors
\begin{equation}
\begin{aligned}
    U_k&=\begin{pmatrix}
        u_k & v_k \\ v_k & u_k
    \end{pmatrix}\\
    u_k &= \frac{\varepsilon_k + z_k}{\sqrt{2\varepsilon_k (\varepsilon_k + z_k)}}\\
    v_k &= \frac{i y_k}{\sqrt{2\varepsilon_k (\varepsilon_k + z_k)}}\\
    u^2_k - v^2_k &=1.
\end{aligned}\label{eq:eigvecs}
\end{equation}

Because the present Hamiltonian is non-Hermitian, this matrix is non-unitary and it does not constitute a canonical transformation. Instead, it defines new creation and annihilation operators that create and annihilate left and right eigenvectors. See for example Ref.~\cite{PhysRevLett.123.123601} for consequences of similar kind in non-Hermitian BCS theory.

The transformation thus defines two new quasiparticle excitations $\gamma_k$ and $\bar{\gamma}_k$ such that $\psi^\dagger_k U_k= \begin{pmatrix}
    \bar{\gamma}_k & \gamma_{-k} 
\end{pmatrix}$ and
$U^{-1}_k \psi_k = \begin{pmatrix}
    \gamma_k \\ \bar{\gamma}_{-k} 
\end{pmatrix}$. We note that $\bar{\gamma}_k\neq \gamma^{\dagger}_k$, but that they still satisfy $\{\bar{\gamma}_k,\gamma_{q}\}=\delta_{k,q}$.
Finally, 
\begin{equation}
\begin{aligned}
    \mathcal{H} &= i\kappa M + 2\sum_{k\in BZ/2} \varepsilon_k (\bar{\gamma}_k \gamma_k + \bar{\gamma}_{-k} \gamma_{-k} - 1)\\
    &= i\kappa M + 2\sum_{k\in BZ} \varepsilon_k (\bar{\gamma}_k \gamma_k - \frac{1}{2}).
\end{aligned}\label{eq:PBCHf}
\end{equation}

Note that Eq.~\eqref{eq:eigvecs} and the diagonalization is not well defined where $u_k=v_k$, where the two eigenvectors would be parallel. This defines exceptional points (EPs). It occurs when $\varepsilon_k=0$, which holds for $\kappa=J$ and $k=(2m+1)\pi/2$ ($m \in \mathds{Z}$), where there is indeed behaviour characteristic of a $PT$-symmetric model - the spectrum goes from complex (imaginary in this case) to purely real, and this transition happens at the EP, as can be seen on Fig.~\ref{fig:PBCspec}.

\subsection{Non-Hermitian TFIM with OBC and edge deformation}

After finding the bulk spectrum, we turn to the actual problem with the open boundary conditions (OBC) and the different possible configurations of the transverse field at the boundaries, Eq.~\eqref{eq:nh-ising}. 

To solve this, we follow the Lieb-Schultz-Mattis method~\cite{Lieb1961Dec,Pfeuty1970Mar,PhysRevLett.25.443} used to solve the TFIM with OBC. 
Ref.~\cite{Mahyaeh2018Apr} is a more modern paper where it has been used for more general Kitaev models. Further, the TFIM with aligned or anti-aligned longitudinal boundary fields has also been studied~\cite{Campostrini_2015} using the same methods. Here we adapt it to the non-Hermitian case.

We perform a Jordan-Wigner transform of~\eqref{eq:nh-ising} and write the model in terms of matrices $A$ and $B$:

\begin{equation}
\begin{aligned}
    \mathcal{H}_{\text{eff}}&=J\sum_{l=1}^{M} (a_l^\dagger-a_l)(a_{l+1}^\dagger+a_{l+1}) + i\kappa\sum_{l=2}^{M} (1-2a_l^\dagger a_l) \\ &+ i\kappa \zeta_L (1-2a_0^\dagger a_0) + i\kappa \zeta_R (1-2a_{M+1}^\dagger a_{M+1})\\ &= i\kappa (M+\zeta_L+\zeta_R-1) \\ &+ 2 \sum_{i,j=1}^{M+1} \left[ a_i^\dagger A_{ij} a_j + \frac{1}{2} a_i^\dagger B_{ij} a_j^\dagger + \frac{1}{2} a_i B^\dagger_{ij} a_j \right]\\ &\equiv i\kappa(M+\zeta_L+\zeta_R-1) + 2 H.
\end{aligned}
\end{equation}

The matrices are (we define $\mu\equiv -i\kappa $)
\begin{equation}
\begin{aligned}
    A&=\frac{1}{2} \begin{pmatrix}
2\mu\zeta_L & J &    &  &    &    \\
J & 2\mu       & J   &     &  & \\
      &       J  & 2\mu & J &  & \\
      &         &     \ddots      &     \ddots      &  \ddots & \\
      &     &    & J & 2\mu & J \\
      &     &    &           & J & 2\mu\zeta_R
    \end{pmatrix}\\
    B&=\frac{1}{2} \begin{pmatrix}
0 & J &    &  &    &    \\
-J & 0       & J  &     &  & \\
      &       -J  & 0 & J &  & \\
      &         &     \ddots      &     \ddots      &  \ddots & \\
      &     &    & -J & 0 & J \\
      &     &    &           & -J & 0
    \end{pmatrix}.
\end{aligned}
\end{equation}

Now, analogously to the PBC case, we define new quasiparticles $\gamma_\alpha = \sum_{i=1}^{M+1} (g_{\alpha i} a_i^\dagger + h_{\alpha i}a_i )$, with $\alpha=1,..., M+1$. In the Hermitian limit this would be a canonical Bogoliubov transform. 
It puts the Hamiltonian into the same form as~\eqref{eq:PBCHf}, except that $k$ is not the momentum anymore and its possible values $k_\alpha$ need to be determined. (We label the quasiparticles by $k$ instead of $\alpha$ in the following.) Supposing $\{\bar{\gamma}_k,\gamma_{q}\}=\delta_{k,q}$ will also hold here, we can write $[H,\gamma_k]=-\varepsilon_k \gamma_k$. Explicitly writing out the transform and using the fact that $A$ is symmetric and $B$ is real, leads to 
\begin{align}
    \sum_i (g_{\alpha i} A_{ij} - h_{\alpha i} B_{ij}) &= -\varepsilon_{k_\alpha} g_{\alpha j} \\
    \sum_i (g_{\alpha i} B_{ij} - h_{\alpha i} A_{ij}) &= -\varepsilon_{k_\alpha} h_{\alpha j}.
\end{align}
Summing them and subtracting, then defining $\phi_{\alpha i} = h_{\alpha i} + g_{\alpha i}$ and $\psi_{\alpha i} = h_{\alpha i} - g_{\alpha i}$ yields 
\begin{align}
    \phi_{\alpha} (A-B) &= \varepsilon_{k_\alpha} \psi_{\alpha} \\
    \psi_{\alpha} (A+B) &= \varepsilon_{k_\alpha} \phi_{\alpha}.
\end{align}
Finally multiplying them together gives decoupled eigenvalue equations that we need to solve (one of them is sufficient) to determine the energy levels end eigenstates of the system
\begin{align}
    \phi_{\alpha} (A-B)(A+B) &= \varepsilon^2_{k_\alpha} \phi_{\alpha} \label{eq:phialpha}\\
    \psi_{\alpha} (A+B)(A-B) &= \varepsilon^2_{k_\alpha} \psi_{\alpha}.
\end{align}

We now solve $\psi_\alpha C = \varepsilon^2_{k_\alpha} \psi_\alpha$, where $C=(A+B)(A-B)$, the matrix to diagonalize, is:

\begin{equation}
    C=\begin{pmatrix}
 J^2+\mu^2\zeta_L^2 & \mu J & &  & \\
 \mu J  & J^2+\mu^2 & \mu J &  & \\
 & \ddots & \ddots &  \ddots & \\
 & & \mu J & J^2+\mu^2 & \mu J\zeta_R \\
 & & & \mu J\zeta_R & \mu^2\zeta_R^2
    \end{pmatrix}.\label{eq:matrixC}
\end{equation}

First, we consider another matrix with the same bulk, of size $n\cross n$  
\begin{equation}
    D_n=\begin{pmatrix}
J^2+\mu^2 & \mu J &  &  &  \\
\mu J & J^2+\mu^2  & \mu J &  & \\ & \ddots &  \ddots      &  \ddots & \\     &    & \mu J & J^2+\mu^2 & \mu J \\
      &     &  & \mu J & J^2+\mu^2
    \end{pmatrix}.
\end{equation}

To find its spectrum, we can rewrite the secular equation recursively for system size. As it has the same bulk as the PBC model, we choose an ansatz of the form of~\eqref{eq:PBCenergy} (squared), but with unknown $k$:
$\lambda_k=J^2+\mu^2-2\mu J\cos k$.
Hence
\begin{equation}
    d_n=\det(D_n-\lambda_k)=2J\mu\cos k d_{n-1} - J^2 \mu^2 d_{n-2}.
\end{equation}
This recursive relation (see Ref.~\cite{Campostrini_2015}) is solved by assuming a solution of the form $d_n=x^n$, which leads to a characteristic equation, whose roots are the possible values of $x$, such that
\begin{equation}
    d_n=\alpha (\mu J)^n e^{ikn} + \beta (\mu J)^n e^{-ikn},
\end{equation}
with $\alpha$ and $\beta$ arbitrary constants. By requiring $d_1=2\mu J \cos k$ and $d_2=(\mu J)^2(4\cos^2k-1)$, these constants need to be $\alpha=e^{ik}/(2i\sin k)$ and $\beta=\alpha^*$, which leads to the following solution
\begin{equation}
    d_n=(\mu J)^n\frac{\sin(k(n+1))}{\sin k}.
\end{equation}
Imposing that it is equal to 0 allows to find all possible values of $k$ and therefore all eigenvalues for the $n\cross n$ matrix.

We now observe that the matrix $C$ is 
\begin{equation}
    C_n=\begin{pmatrix}
  J^2+\mu^2\zeta_L^2 & \begin{matrix} \mu J & & & \end{matrix} &  \\
  \begin{matrix} \mu J \\ \\ \end{matrix}  &
  \begin{pmatrix}
  \hspace*{-\arraycolsep}
  \phantom{x_{11}} & \phantom{x_{11}} & \phantom{x_{11}}
  \hspace*{-\arraycolsep}
  \\
  & \raisebox{-0.2\height}[0pt][0pt]{\Large$D_{n-2}$} & \\
  & &
  \end{pmatrix} & \begin{matrix} \\ \\ \mu J\zeta_R \end{matrix} \\
   & \begin{matrix} & & & \mu J\zeta_R  \end{matrix} & \mu^2\zeta_R^2
\end{pmatrix}.
\end{equation}
Therefore, taking the same ansatz for the eigenvalues and after some calculations, the secular equation can be written in terms of $d_n$ 
\begin{equation}
\begin{aligned}
    c_n=&\det(C_n-\lambda_k)=\left[\mu^2\zeta_L^2\zeta_R^2-\lambda_k\right]d_{n-1} \\+& \left[ \mu^2 J^2(1-\zeta_L^2-\zeta_R^2) \right. \\+& \left. \mu^2 (\zeta_L^2-1)(\zeta_R^2-1)(\mu^2-2\mu J\cos k) \right] d_{n-2}.
\end{aligned} \label{eq:OBCsecular}
\end{equation}

Now, in order to find all possible values of $k$ and thus the eigenvalues $\lambda_k$ for our model, we solve $c_{M+1}\overset{!}{=}0$ for the different boundary configurations set out by $\zeta_{L/R}$:

\begin{equation}
    \begin{aligned}
        &\left[\mu^2\zeta_L^2\zeta_R^2-\mu^2-J^2+2\mu J \cos k\right](\mu J)\frac{\sin(k(M+1))}{\sin k} \\&= -  \left[ \mu^2 J^2(1-\zeta_L^2-\zeta_R^2) \right. \\+& \left. \mu^2 (\zeta_L^2-1)(\zeta_R^2-1)(\mu^2-2\mu J\cos k) \right] \frac{\sin(kM)}{\sin k} .
    \end{aligned} \label{eq:OBCsecular_end}
\end{equation}

We can now consider three different possibilities for $\zeta_{L/R}$ in Eq.~\eqref{eq:OBCsecular_end} and find the corresponding eigenvalues. Then, in Sec.~\ref{appendix:eigenvecs}, we discuss the form of the eigenvectors, and in particular of the zero modes.

\subsubsection{OBC with one edge field ($\zeta_{L}=1,\zeta_{R}=0$ or $\zeta_{L}=0,\zeta_{R}=1$)}

This case is the simplest to solve, as Eq.~\eqref{eq:OBCsecular_end} simplifies to
\begin{equation}
    \lambda_k\frac{\sin(k(M+1))}{\sin k}= 0 . \label{eq:OBCone}
\end{equation}
There is a trivial $\lambda_k=0$ eigenvalue.
Another set of solutions is given by 
\begin{equation}
    k_{\alpha}=\frac{\alpha \pi}{M+1},
\end{equation}
where $\alpha \in \mathds{Z}\setminus (M+1)\mathds{Z}$, that is all integers except multiples of $M+1$.
We only need the values of $k$ contained in the interval defined by $\Re{k} \in \left[0,\pi \right)$ in order to get all the $M+1$ solutions to the eigenvalue problem.

\subsubsection{OBC with both edge fields \texorpdfstring{($\zeta_{L}=\zeta_{R}=1$)}{}}

\begin{figure}[!ht] 
	\centering
	\includegraphics{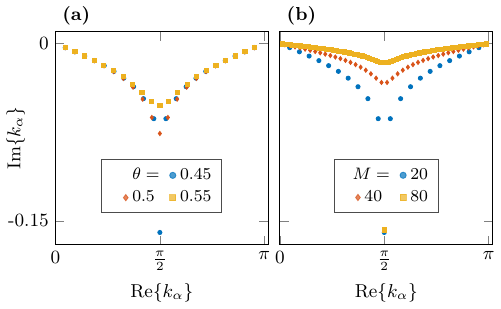}
    \caption{Numerical solution of~\eqref{eq:OBCboth} for the complex "momenta" $k_{\alpha}$. (a) Shown for different parameters set by $\theta$, such that $J=\cos(\theta \pi/2)$ and $\kappa=\sin(\theta \pi/2)$, for system size $M=20$. (b) Shown for different system sizes $M$, with $\theta=0.45$.}
	\label{fig:ZIZ_appendix}
\end{figure}

In this case, Eq.~\eqref{eq:OBCsecular_end} becomes: 
\begin{equation}
    \frac{\sin(k(M+1))}{\sin k}= \frac{\mu}{J} \frac{\sin(k(M+2))}{\sin k} . \label{eq:OBCboth}
\end{equation}
This is the equation corresponding to a regular TFIM with OBC~\cite{Pfeuty1970Mar}, but here with an imaginary parameter $\mu$. 
Using Chebyshev polynomials of the second type, $U_n(x)$, which are defined by $U_n(\cos \theta) = \sin((n+1)\theta) /\sin \theta $, Eq.~\eqref{eq:OBCboth} can be written as a univariate polynomial equation for $\cos k$:
\begin{equation}
    \frac{\mu}{J} U_{M+1}(\cos k) - U_{M}(\cos k) = 0 . \label{eq:OBCboth_2}
\end{equation}
In this case, the polynomial is of order $M+1$, which informs us that we can find $M+1$ roots in the complex plane. The corresponding numerical solutions for $k$ with real parts in the $\left[0,\pi \right)$ interval are displayed in \figref{fig:ZIZ_appendix}[a] for different parameter regimes.

From the appearance of the solutions $k_{\alpha}$, we observe two phases, with a transition around $\kappa=J$, distinguished by the presence or absence of an eigenmode of zero energy. The existence of such a zero mode and of the phase transition can be inferred from the following simple argument. 

Starting from Eq.~\eqref{eq:OBCboth}, ignoring the common denominator, and writing the sines in terms of exponentials, we obtain:
\begin{equation}
    e^{ikM}\left( e^{ik} + i\frac{\kappa}{J} e^{i2k} \right) = e^{-ikM}\left( e^{-ik} + i\frac{\kappa}{J} e^{-i2k} \right). \label{eq:zeromode_1}
\end{equation}
Assuming that we look for solutions $k$ with strictly positive imaginary part, we take the thermodynamic limit $M\rightarrow \infty$ which makes the left-hand side of \eqref{eq:zeromode_1} vanish and we are left with: $ e^{ik} = -i\kappa /J$. This is satisfied by 
\begin{equation}
    k_{zm}=-\pi/2 - i \log(\kappa/J) + 2\pi\beta, \label{eq:zm1}
\end{equation}
with $\beta \in \mathds{Z}$. However, this solution only exists for $\kappa < J$, since we assumed the imaginary part of $k$ is positive.

Similarly, if we look for such solutions in the lower half of the complex plane, that is by assuming the imaginary part of $k$ to be strictly negative, the right-hand side of \eqref{eq:zeromode_1} vanishes as $M\rightarrow \infty$ and we get: $ e^{-ik} = -i\kappa /J$. This is true for 
\begin{equation}
    k_{zm}=\pi/2 + i \log(\kappa/J) + 2\pi\beta, \label{eq:zm2}
\end{equation} 
with $\beta \in \mathds{Z}$. Again, this solution can only exist for $\kappa < J$, since we assumed the imaginary part of $k$ is negative.

\begin{figure}[!ht] 
	\centering
	\includegraphics{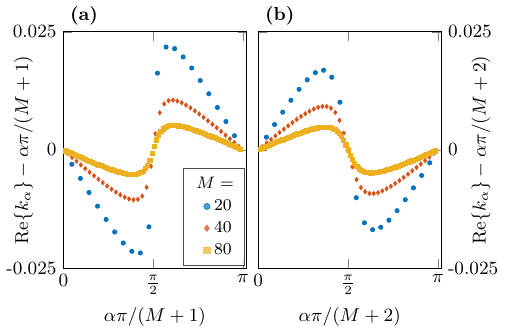}
	\caption{Numerical solution of~\eqref{eq:OBCboth}. Displacement of the real part of $k_\alpha$ as compared to $\alpha\pi/(M+1)$ or $\alpha\pi/(M+2)$. (a) Below the transition, at $\theta=0.45$ (such that $J=\cos(\theta \pi/2)$ and $\kappa=\sin(\theta \pi/2)$). Here the zero mode is excluded from the set of solutions for $k_\alpha$. (b) Above the transition, at $\theta=0.55$.}
	\label{fig:ZIZ_appendix_re}
\end{figure}

In both cases, the eigenmodes that we found have zero energy, as given by Eq.~\eqref{eq:PBCenergy}, and exist only in the $\kappa < J$ regime. As $M$ tends to infinity, these zero modes are the only ones that have complex momentum -- all the other ones tend to equally spaced solutions with real momenta. 
This is illustrated on \figref{fig:ZIZ_appendix}[b]. Moreover, \figref{fig:ZIZ_appendix_re}[a] shows that below the transition, as $M$ is increased, the solutions tend to $k_\alpha=\alpha\pi/(M+1)$. On the other hand, \figref{fig:ZIZ_appendix_re}[b] shows that above the transition, they tend to $k_\alpha=\alpha\pi/(M+2)$.  

This proves the existence of the two phases. The form of the zero eigenmode is further described in Sec.~\ref{appendix:eigenvecs}.  

\subsubsection{OBC without edge fields \texorpdfstring{ ($\zeta_{L}=\zeta_{R}=0$)}{}}

In this last case, Eq.~\eqref{eq:OBCsecular_end} turns into:
\begin{equation}
    \lambda_k \frac{\sin(k(M+1))}{\sin k}= \lambda_k \frac{\mu}{J} \frac{\sin(kM)}{\sin k} . \label{eq:OBC}
\end{equation}
It can again be written as a polynomial equation using Chebyshev polynomials:
\begin{equation}
    \lambda_k U_M(\cos k) - \lambda_k \frac{\mu}{J} U_{M-1}(\cos k) = 0 . \label{eq:OBC_2}
\end{equation}
Besides the trivial zero eigenvalue $\lambda_k=0$, another $M$ complex solutions for $\cos k$ are given by the roots of the $M^{\text{th}}$ order polynomial, which can be obtained numerically. 

If we apply the same reasoning as for the previous case, we find that in the thermodynamic limit there are zero modes only in the $\kappa > J$ phase. 

This case cannot actually occur in a fragment of our model, because missing the transverse field at both ends implies the boundaries of the fragment need to be the boundaries of the whole system. Yet, in our model, odd and even sublattices are decoupled, and one such sublattice cannot contain both ends of a lattice, which we assumed to have an even number of sites.

\subsubsection{Eigenvectors} \label{appendix:eigenvecs}

An eigenvector $\psi_\alpha = (u_1,u_2,...,u_{M+1})$ of matrix $C$ in Eq.~\eqref{eq:matrixC} needs to satisfy:
\begin{align}
u_2 &= \left( \frac{\mu}{J}\left(1- \zeta^2_L\right) - 2\cos k_\alpha  \right) u_1,\nonumber\\
0 &= u_{j-1} + 2\cos k_\alpha u_{j} + u_{j+1}, \;\;\; (j=2,...,M-1)\nonumber\\
0 &= u_{M-1} + 2\cos k_\alpha u_{M} + \zeta_R u_{M+1},\nonumber\\
0 &= \zeta_R u_{M} + \left( \frac{\mu}{J}\left( \zeta^2_R -1 \right) - \frac{J}{\mu} + 2\cos k_\alpha  \right) u_{M+1}
.\label{eq:syst}
\end{align}
The recurrence relation for $u_j$ has the following solution:
\begin{equation*}
    u_j=(-1)^j\left( c_1 e^{ik_\alpha j} + c_2 e^{-ik_\alpha j} \right),
\end{equation*}
with $c_1,c_2$ constants to be determined.

We now look at the zero modes \eqref{eq:zm1} and \eqref{eq:zm2} discovered in the case $\zeta_L=\zeta_R=1$. The first equation of \eqref{eq:syst} with $\zeta_L=1$ leads to $c_1=-c_2$, so $u_j$ becomes
\begin{equation}
    u_j=(-1)^j c \sin(k_\alpha j),\label{eq:eigenmodes_psi}
\end{equation}
with $c$ a normalization constant. For any $k_\alpha$ with a nonzero imaginary part, the wave function is localized towards the right edge of the system.

Similarly, the $\phi_\alpha=(v_1,...,v_{M+1})$ eigenvectors from Eq.~\eqref{eq:phialpha} are given by
\begin{equation}
    v_j=(-1)^j d \sin(k_\alpha (j-M-2)),\label{eq:eigenmodes_phi}
\end{equation}
with $d$ some normalization constant. They are found to localize near the left edge of the system if the corresponding $k_\alpha$ has a nonzero imaginary part.

\begin{figure}[!ht] 
	\centering
	\includegraphics{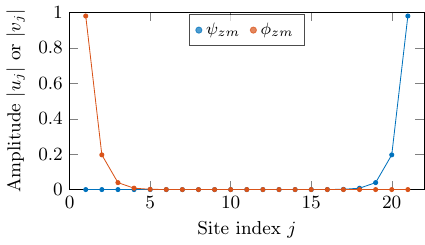}
	\caption{Localization of the zero modes \eqref{eq:zm2} on the edges. The plot shows the amplitude of the components of the modes $\psi_{zm}$ and $\phi_{zm}$, given by \eqref{eq:eigenmodes_psi} and \eqref{eq:eigenmodes_phi} with $k_\alpha = k_{zm}$. The system size is $M=20$ and $\kappa/J=0.2$.}
	\label{fig:zeromodes}
\end{figure}

In particular, in the thermodynamic limit, the zero modes are localized at the edges of the system and have exponentially decaying tails in the bulk, as illustrated on Fig.~\ref{fig:zeromodes}.

\bibliography{Bibliography}

\end{document}